\documentclass[a4paper,times]{cjeart}
\pdfoutput=1
\usepackage{amsthm} 

\graphicspath{{figures/}}
\usepackage{graphicx}

\usepackage{amssymb}
\usepackage{balance}
\usepackage{booktabs}

\usepackage[active]{srcltx}
\usepackage{caption}
\usepackage{latexsym,bm}
\usepackage{multirow}
\usepackage{epsfig} 
\usepackage{extarrows}
\usepackage{chemarrow}

\usepackage{amsmath} 
\usepackage{algorithmic}
\usepackage[linesnumbered,ruled,vlined]{algorithm2e}
\usepackage{flushend}

\usepackage{subfigure}
\usepackage{bm}
\usepackage{marvosym}
\usepackage{color}
\usepackage{epstopdf}
\definecolor{ColorBlue}{RGB}{51,102,205}

\volumeyear{2022}
\volumenumber{04}
\issuenumber{04}
\startpage{1}
\DOI{123.123.123.123}
\journaltype{Research Article}

\title[Resource-aware Probability-based Collaborative Odor Source Localization Using Multiple UAVs]
{Resource-aware Probability-based Collaborative Odor Source Localization Using Multiple UAVs}

\author{%
Shan Wang\affilnums{1,2},
Sheng Sun\affilnums{1,2},
Min Liu\affilnums{1,2,3},
Bo Gao\affilnums{4}, and \\
Yuwei Wang\affilnums{1}
}

\affiliation{%
\affilnum{1}Institute of Computing Technology, Chinese Academy of Sciences, Beijing 100190, China\\
\affilnum{2}University of Chinese Academy of Sciences, China \\
\affilnum{3}Zhongguancun Laboratory, China \\
\affilnum{4}School of Computer and Information Technology, Beijing Jiaotong University, Beijing 100044, China
}

\abstract{%
Benefitting from UAVs' characteristics of flexible deployment and controllable movement in 3D space, odor source localization with multiple UAVs has been a hot research area in recent years. Considering the limited resources and insufficient battery capacities of UAVs, it is necessary to fast locate the odor source with low-complexity computation and minimal interaction under complicated environmental states.
To this end, we propose a multi-UAV collaboration based odor source localization (\textit{MUC-OSL}) method,
where source estimation and UAV navigation are iteratively performed, aiming to accelerate the searching process and reduce the resource consumption of UAVs.
Specifically, in the source estimation phase,
we present a collaborative particle filter algorithm on the basis of UAVs' cognitive difference and Gaussian fitting to improve source estimation accuracy.
In the following navigation phase, an adaptive path planning algorithm is designed based on Partially Observable Markov Decision Process (POMDP) to distributedly determine the subsequent flying direction and moving steps of each UAV.
The results of experiments conducted on two simulation platforms demonstrate that \textit{MUC-OSL} outperforms existing efforts in terms of mean search time and success rate, and effectively reduces the resource consumption of UAVs.
}


\begin{document}

\maketitle

\section{Introduction}
\label{sec1}
Odor source localization (OSL) for environmental emergency and atmospheric control has received considerable interest and become a hot research area
in recent years\cite{wang2022robotic,zhao2022deep,RAL2021infoDriven}.
Finding the source of gaseous chemicals released in the air is vital in many situations, such as rescue and search, air quality monitoring, and security assurance\cite{furukawa2006recursive,li2011odor, MUDULI201848}.
Traditional solutions utilize static sensors and mobile robots to collect data and locate the source,
but they are scope-limited and invalid for high-altitude sources.
To this end, Unmanned Aerial Vehicles (UAVs) carrying appropriate odor monitoring sensors are applied in the odor source localization field due to their advantages of flexible deployment and controllable mobility in 3D space \cite{rossi2014gas, kersnovski2017uav, JNCA2020}.
Recently, this scheme is extended from single UAV to a team of UAVs to locate odor source in a cooperative behavior,
which can greatly shorten the search time and improve the localization fault tolerance, becoming the mainstream in the future \cite{duisterhof2021sniffy, access19multiUAV}.

The methods of odor source localization can be mainly classified into gradient-based, bio-inspired, and probability-based methods  \cite{Inffus17review}.
Gradient-based method climbs the concentration gradient to approach the source,
which is
inefficient in a turbulence-dominated environment with unstable and intermittent airflow.
Bio-inspired odor source localization takes inspiration from living organisms such as moths, dogs, bacteria, etc \cite{kowadlo2008robot}, which leverages perceived odor plume and wind information to guide movement.
However, it cannot take into account dynamic disturbances in realistic environment such as variable wind field that observably affect searching effectiveness of the source localization.
Probability-based source localization method formulates the process of source localization as the update of probability map.
The update is iteratively performed using Bayesian estimation after obtaining gas measurement.
This method can adapt to various environmental conditions and explore the probable impact of environmental factors such as sensor noise and unsteady atmospheric conditions on source localization.
Hence, compared to other methods, it is more efficient and reliable to locate odor source in the real world.
Concretely, it involves two primary phases: estimation and navigation \cite{rahbar2019algorithm}.
The former focuses on inferring source probability distribution based on sensor measurements, and the latter aims to plan a path toward the location to get the most information based on the results derived from the estimation phase.


However, existing probability-based source localization approaches are only applicable to 2D space \cite{07infotaxis, 19TRO,  hutchinson2018information, inffus18entrotaxis}.
In these works, particle filter as an approximation method of probability distribution is gradually applied to estimate the source \cite{li2011odor}.
The particle weights and positions are iteratively updated to infer source parameters and draw near source location in the source searching process.
In collaborative localization based on information sharing, neighbors share each other's particles states to extend the perception range and increase estimation accuracy.
To improve the approximation accuracy of particle filter, a large number of particles will be placed to infer source parameters, which will lead to a large amount of data transfer when sharing information and require more computation time for updating particles in each iteration.
On the basis of real-time inferences,
information-driven path planning methods are commonly-used in the navigation phase to determine the optimal control action,
which mainly focuses on how to decide the next moving direction to obtain the maximum information under the fixed step size.
However, maintaining a fixed step size throughout the searching process is time-consuming and ineffective, particularly in the huge 3D search space \cite{Inffus17review}.
Therefore, existing methods are unable to be directly applied to the multi-UAV odor source localization scenario in 3D space, since limited resources and insufficient battery capacities of UAVs cannot afford high resource requirements and large energy consumption induced by real-time probability inference and multi-UAV collaboration.
\textbf{To the best of our knowledge, our work is the first to study the probability-based odor source localization using multiple UAVs in 3D space.}

There are three major challenges to be tackled in the multi-UAV odor source localization scenario in the 3D space.
First, it naturally incurs an increase in computation to infer and update source probability map with a large number of particles in the high-dimensional and large-scale searching space, making it challenging for resource-constrained UAVs.
Second, it is far from trivial to timely and efficiently determine multiple UAVs' moving paths because of the increased action space and the joint decision making, inevitably increasing the likelihood of erroneous moving and slowing down the searching speed.
Moreover, UAVs with limited energy is hurdled to cope with the heavy communication overhead and energy consumption, caused by frequent interactions between multi-UAV to share a large amount of local searching information with each other to improve the ability of environment perception and decision efficiency.
Thus, a probability-based method with low resource consumption and fast searching pace is required to efficiently locate odor source with multi-UAV collaboration.

To this end, we propose multi-UAV collaboration based resource-aware odor source localization algorithm (\textit{MUC-OSL}) involving collaborative particle filter and adaptive path planning to achieve a fast and low-cost 3D source localization.
Specifically, we utilize insights of cognitive difference and distribution fitting among multiple UAVs to estimate source position, so as to accelerate particle convergence and cut down resource overhead.
After obtaining source inferences, we adaptively adjust multiple UAVs' moving positions with befitting steps depending on inferred source probability and sequential gas measurements, aiming to avoid the path redundancy and reduce the search time.
Each UAV repetitively performs the preceding steps until source declaration condition is triggered.
The main contributions of this paper are summarized as follows:
\begin{itemize}
\item In the estimation phase, utilizing cognitive differences to selectively fuse information of neighboring UAVs, we propose cooperative particle filtering (\textit{Col-PF}) algorithm to real-timely infer source probability distribution.
    On the basis of adaptively updating the number of particles,     the weight and position of each particle are updated, considering global and local information, to reduce the computational overload and speed up the particle convergence.
    Afterwards, the distribution of particles is fitted before information sharing to avoid the high communication load.


\item In the navigation phase, we model the sequential multi-UAV path planing problem as Partially Observed Markov Decision Process (POMDP), and then propose adaptive path planning (\textit{Adap-PP}) algorithm to adaptively determine multiple UAVs' subsequent positions.
    Under the chosen direction of the maximum decrease in estimation uncertainty,
    the moving steps is adaptively determined in association with the historic measurement information and distance from the UAV to the estimated source, so as to shorten localization time and reduce UAVs' energy consumption.

\item We verify the effectiveness of our proposed algorithm (\textit{MUC-OSL}) in terms of mean search time and success rate through extensive simulations.
    The profound impact of simulation parameters, such as UAVs number and searching area size, and multi-UAV resource consumption also be considered in the simulations.
    Simulation results show that \textit{MUC-OSL} algorithm can fast locate odor source with competitive success rate while reducing resource consumption.

\end{itemize}

The remainder of the paper is organized as follows.
Section \uppercase\expandafter{\romannumeral2}
presents the related work on the issue of odor source localization.
Section \uppercase\expandafter{\romannumeral3}
describes the system model and whole system framework.
Section \uppercase\expandafter{\romannumeral4}
introduces the collaborative particle filter algorithm to estimate source parameters. 
Section \uppercase\expandafter{\romannumeral5}
elaborates the multi-UAV adaptive path planning algorithm. 
Section \uppercase\expandafter{\romannumeral6}
presents the performance evaluation results.
Finally, the conclusion of this paper is drawn in
Section \uppercase\expandafter{\romannumeral7}.

\section{Related Work}
\label{S:2}

There has been growing research interest in odor source localization over the probability-based method.
Infotaxis, a typical probability-based odor source localization strategy, is proposed in \cite{07infotaxis} to search the diffusive source in a complex and turbulent environment.
The probability that each region encloses the source is inferred based on new sensor measurements, and then information-driven path planning is performed to acquire more information about the source.

The estimation of source parameters is enabled by sequentially inferring source probability distribution based on the sensor detections.
In earlier studies\cite{hajieghrary2016multi,karpas2017information}, grid-based probability map is used to locate odor source through iteratively updating each grid's occupancy probability of enclosing the odor source. 
The grid with the maximum probability value corresponds to the most probable source position.
To improve the estimation accuracy, it is necessary to increase the number of grids covering the whole searching area, which will lead to a challenge of high computation cost.
Moreover, multiple robots collaboratively search odor source by exchanging their occupancy grid maps with others, which consumes a large amount of communication resources \cite{karpas2017information}.

In order to reduce the resource consumption of grid-based localization, particle filter is proposed to approximate the probability map of odor source via iteratively updating the particles' positions and weights.
The particle filter based source localization framework is first presented in \cite{li2011odor}.
Then a lot of researches gradually use particle filter to estimate the source parameters \cite{ 19TRO,hutchinson2018information,inffus18entrotaxis, 20-IJRR}.
However, the exchanged information between neighboring robots usually includes the positions and the weights of all particles related to environmental state update in these existing works \cite{Inffus16study,hajieghrary2017information},
which leads to expensive computation and communication burden, especially when using a larger number of particles.
How to reduce the communication and computation consumption for resource-limited UAVs is an essential problem to be addressed.
In this paper, we utilize limited and adaptive particles to estimate the source based on multi-UAV collaboration to avoid the high resource consumption.

After the probability map of the inferred odor source is obtained, mobile robots plan their paths to find the source as soon as possible in the navigation phase.
Information-driven path planning has become the dominant trend, which allows mobile robots to determine their next positions based on real-time information\cite{wei2020informative}.
This process is usually modeled as POMDP, including three main elements: the information state, the action space and the reward function.
The key is to design an appropriate reward function to determine the movement directions of mobile robots.  
Inspired by the maximum sampling theory \cite{sebastiani2000maximum}, Entrotaxis strategy in \cite{inffus18entrotaxis} defines the reward function as the maximum entropy of the expected measurement distribution, and it shortens the average search time compared with the above-mentioned infotaxis strategy.
Then three reward functions based on information theory, including infotaxis reward, infotaxis II reward (removing infotaxis bias), and Bhattacharyya distance, are presented in \cite{Inffus16study}.
What is more, the mutual information between model outputs and sensor measurements is defined as the reward function to guide the robot to move \cite{madankan2014optimal}.
The R\'{e}nyi divergence \cite{ristic2010information}, Kullback-Lieber divergence \cite{hajieghrary2016multi} or its variants can also be used as the reward function in the information-driven odor source localization \cite{jing2021recent}.
However, all reward functions designed in these methods adopt a fixed moving step, which impacts the searching efficiency in the huge 3D space.
In this paper, we dynamically adjust the moving steps based on the real-time states to avoid the path redundancy, so as to reduce the search time and improve the localization efficiency.

\section{System Model and Framework}
\subsection{Measurement Model}
The role of measurement model is to model the chemical plume behavior and capture diffusive particulates in turbulent environment.
For the odor source located at $r_s=[x_s, y_s, z_s]$ with particulate release rate $Q$, the average rate of encountering particulates by a spherical detecting sensor mounted on the UAV $u_i=[x_i, y_i, z_i]$ can be depicted as follows:
\begin{equation}
    R({u_i}|{r_s}) = \frac{{aQ}}{{|{u_i} - {r_s}|}}\exp \left[ {- \frac{{(y_i - {y_s})V}}{{2D}}} \right]\exp \left[ { - \frac{{|{u_i} - {r_s}|}}{\lambda }} \right],
\end{equation}
where
\begin{equation}
    \lambda  = \sqrt {\frac{{D\tau }}{{1 + \frac{{{V^2}\tau }}{{4D}}}}}.
\end{equation}
Thereinto, $\lambda$ denotes the average distance of each particulate from emission to disappearance, which is affected by the isotropic effective diffusivity $D$, the wind speed $V$ and the limited lifetime $\tau$ of particulates released by the odor source.
$a$ denotes the detecting radius of the spherical sensor.
This model is first proposed in \cite{07infotaxis}, and it is a rate-based plume model.


The stochastic process of UAV encountering particulates can be modeled as Poisson distribution \cite{Inffus16study}.
The probability of encountering  $d_i^k$ particulates by UAV $i$ at current position $u_i^k$ at the $k$-th iteration is represented as follows:
\begin{equation}
    p(d_i^k) = \frac{({R({u_i^k}|{r_s})\Delta t})^ {d_i^k} } {{d_i^k!}}\exp ( - R({u_i^k}|{r_s})\Delta t),
\end{equation}
where ${R({u_i^k}|{r_s})\Delta t}$ represents the expectation of the number of encountered particulates during the time interval $\Delta t$.

\subsection{UAV Energy Consumption Model}
A widely-used energy consumption model of UAVs \cite{zhan2020completion, liu2020path, wang2021joint} is utilized to calculate UAVs resource consumption in source localization.
The total energy consumption of each UAV is composed of following three parts:

\noindent \textbf{Movement energy consumption.}
The movement energy consumption $E_{M}$ of the UAV is comprised of flying, hovering and turning energy consumption in 3D space, i.e., $E_{M}=E_f + E_h + E_b$.
Specifically, the flying energy consumption $E_f$ depends on the flying time $ T_f$, which is related to the flying distance $l$ at a given flying speed $v$.
Thus, $E_f$ can be given by
\begin{equation}
  E_f = P_f T_f = P_f l/v ,
\end{equation}
where $P_f$ denotes the flying power of the UAV.

We assume that the required hovering time at each hovering position is fixed, denoted by $t_h$.
The whole hovering energy consumption of the UAV is proportional to the number of hovering points $n_{h}$, which can be given by
\begin{equation}
  E_h = P_h T_h = P_f n_h t_h,
\end{equation}
where $P_h$ denotes the hovering power of the UAV.

The turning operation of UAVs requires to slow down in one direction and then speed up in another direction without rotating the fuselage.
We assume that the energy consumption at each turning is the same despite of different turning angles, and denote $ e_b$ as the energy consumption per turning.
The turning energy consumption $E_b$ of the UAV is calculated as
 \begin{equation}
  E_b = n_{b} e_{b},
\end{equation}
where $n_{b}$ denotes the number of turning points.

\noindent \textbf{Computation energy consumption.}
The computation energy consumption for one UAV to process $N_c$ bits of data can be calculated as follows:
\begin{equation}
E_C  = \gamma_c C f_{c}^2  N_c,
\end{equation}
where $\gamma_c$ is the effective capacitance coefficient of the UAV, $C$ is the number of CPU cycles for computing one bit, and $f_c$ is the CPU frequency.

\noindent \textbf{Communication energy consumption.}
The communication energy consumption $E_{T}$ is quantified by the data amount transmitted from UAV $i$ to neighboring UAVs as follows:
\begin{equation}
E_T  = { P_{T}   N_{d} / r_T  },
\end{equation}
where $P_T$ is the transmitting power of the UAV, $N_{d}$ is the amount of transmitting data to neighboring UAVs, and $r_T$ represents the transmitting rate.

Consequently, the total energy consumption $E$ for the UAV can be represented as:
\begin{equation}
E = E_M + E_{C} + E_{T},
\end{equation}
where $E$ should satisfy $E \le E_{max}$, and $E_{max}$ denotes the maximum energy of the UAV.

\subsection{Cognitive Difference Model Between UAVs}

The information fusion of multiple UAVs can reduce the measurement deviance of single UAV, caused by environmental noises and measuring errors, and expand the perception range of environment to improve the accuracy of source estimation and reduce the search time.
We adopt cognitive difference to represent the divergences in the corresponding probability distribution of source inference among UAVs, and utilize Kullback-Leibler divergence to characterize cognitive difference \cite{song2019multi}.
The Kullback-Leibler divergence between UAV $i$ and neighboring UAV $j$ can be represented as $D_{KL}(P_i||P_j) = \sum{P_i (x)log{ \frac{P_i (x)}{P_j (x)}}}$, where $P_i$ and $P_j$ denote the probability distributions of source inference by UAV $i$ and UAV $j$ respectively.

Then confidence factor $\beta$ is proposed to measure the influence of neighboring UAVs' information on this UAV to infer the source position based on cognitive difference.
$\beta _{ij}^k$ characterizes the influence weight of neighboring UAV $j$'s information on the source inference of UAV $i$ at the $k$-th iteration, which can be represented as follows:
\begin{equation}
\beta _{ij}^k = \exp [ - {D_{KL}}(P_i^{k - 1}||P_j^{k - 1})],\;\;\;i \in \mathcal{H}, j \in \mathcal{H}_i,
\label{belta}
\end{equation}
where the exponential function is used to map the cognitive difference into $(0,1]$ interval, i.e., $0 < \beta _{ij}^k \le 1$.
The confidence factor $\beta _{ij}^k $ correlates negatively with
the cognitive difference.
That is, if the difference of two probability distributions related to source inference between UAV $i$ and UAV $j$ is larger, the value of $\beta _{ij}^k$ is smaller.
Specifically, $\beta_{ij}^k\approx 1$ corresponds to the case that
the probability maps of two UAVs are extremely similar, and UAV $i$ completely accepts the information from UAV $j$.
The other extreme case is $\beta _{ij}^k\ \approx\ 0$, where two UAVs' divergence is relatively far and they independently search the source position based on their own information.

\subsection{System Framework}

\begin{figure}[h]
  \centering
  \includegraphics[width=0.97\linewidth]{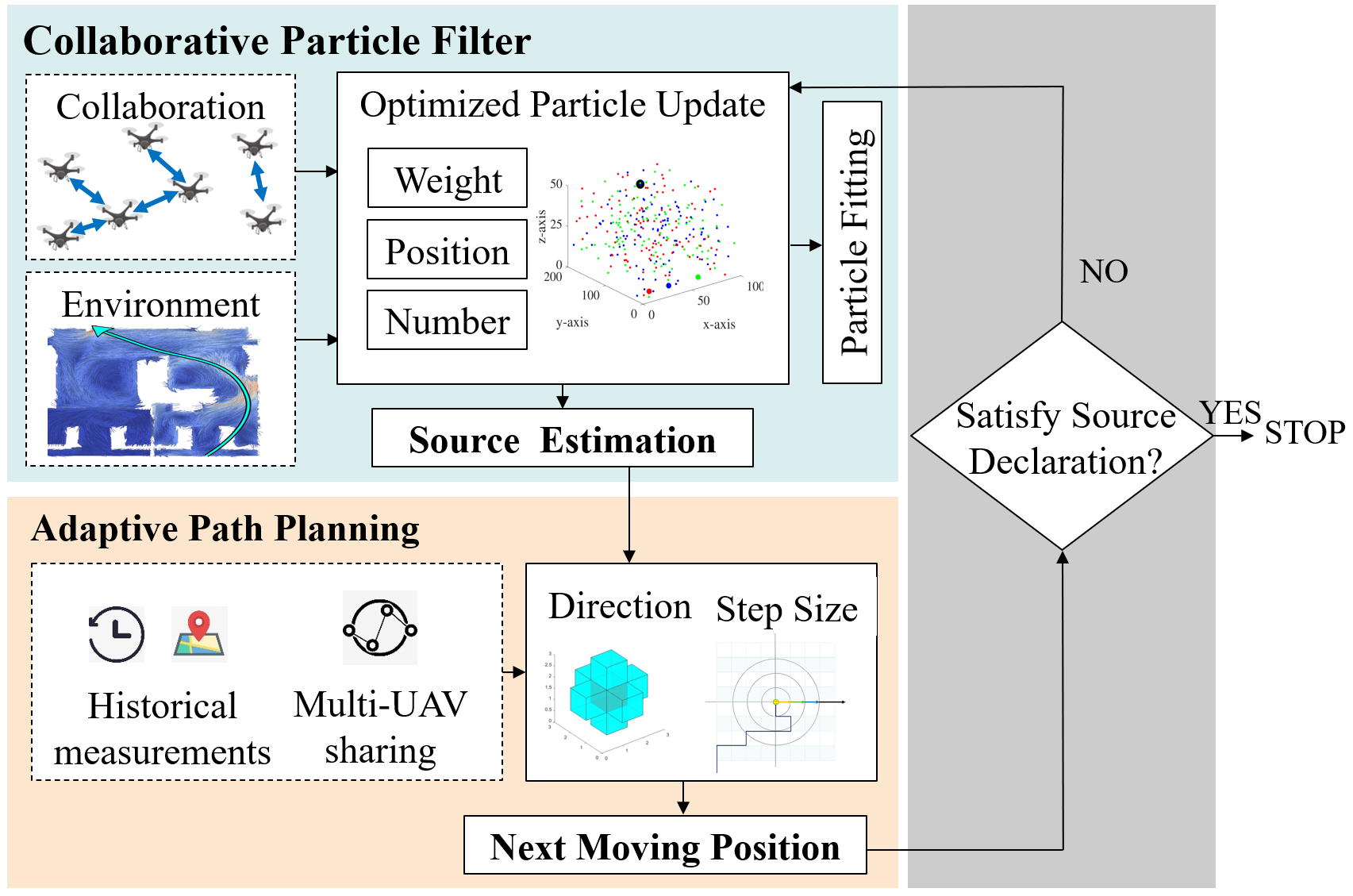}
  \caption{System framework of \textit{MUC-OSL}.}
  \label{framework}
\end{figure}

The system framework of multi-UAV collaboration based odor source localization is shown in Fig. \ref{framework}, consisting of three main components:

\noindent \textbf{Source Estimation.}
This component leverages the collaborative particle filter approach to infer the source parameters based on multi-UAV cognitive difference and external environment states, including  particle position update, particle weight update, particles number update, and particle fitting strategy four main steps, as we will elaborate in Sect. 4. 


\noindent \textbf{Path Navigation.}
This component adaptively determines the next position of each UAV in each iteration, including flying direction (Sect. 5.1) and moving steps (Sect. 5.2), based on historical information, collaboration states, and current estimated results.

\noindent \textbf{Source Declaration.}
After flying to the new position, each UAV implements the source declaration strategy to determine whether to stop source searching.
The odor source is considered found when the uncertainty on the source parameter estimation goes below a certain threshold, and more details can be referred in \cite{rahbar2019algorithm}.


\section{Estimation: Multi-UAV Collaborative Particle Filter Algorithm}
In this section, we mainly focus on source parameter estimation using the proposed collaborative particle filter algorithm in the probability-based odor source localization framework.

\subsection{Optimized Particle Filter}

In the multi-UAV based source localization scenario,
multiple UAVs exchange their real-time information to their neighbors within the communication radius, aiming to expand the perception range and improve the estimation performance.
Source parameters to be estimated mainly refer to source location in this paper.
We adopt Bayesian framework to estimate source parameters.
Let $r^k_{i,s}$ denote the estimated source position by UAV $i$ at the $k$-th iteration.
${d_i^{1:k-1}} = [{d_i^1},{d_i^2}, \cdots ,{d_i^{k-1}}]$ represents historical sequential measurements by the gas-sensitive sensor equipped on UAV $i$ during the iteration period $[1,k-1]$.
The measurement set of UAV $i$'s neighbors under the multi-UAV collaboration can be represented as $\textbf{d}_i^k = [d_1^k, d_2^k,\cdots, d_{\mathcal{H}_i}^k]$, where $d_j^k$ is the measurement of neighboring UAV $j$.
Thus, the likelihood for UAV $i$ is $p(\textbf{d}^k_i|r _{i,s}^k)$ in the multi-UAV scenario, and it can be decomposed as $p(\textbf{d}^k_i|r _{i,s}^k) = \prod\nolimits_{j = 1}^{\mathcal{H}_i} {{{p(d_j^k|r_{i,s}^k)}}}$
since the observations of multiple UAVs are conditionally independent.
The posterior PDF estimated by UAV $i$ under multi-UAV collaboration is updated according to the Bayes rule as follows:
\begin{equation}
    p({r_{i,s}^k}|{d^{1:k}_i}) = \frac{{p(r_{i,s}^k|d^{1:k - 1}_i)p(d_i^k|r_{i,s}^k)\prod\nolimits_{j = 1}^{\mathcal{H}_i} {{{[p(d_j^k|r_{i,s}^k)]}^{\beta _{ij}^k}}} }}{{p({d_i^k}|{d_i^{1:k - 1}})}},
\end{equation}
where
$\prod\nolimits_{j = 1}^{{\mathcal{H}_i}} {{{[p(d_j^k|r_{j,s}^k)]}^{\beta _{ij}^k}}} $ represents the contribution of all neighbors' information on estimating source location for UAV $i$ at the $k$-th iteration, considering that each UAV has different sampling locations and measurements.

Due to the complicated calculation in Bayesian inference as above, particle filter is used to approximate the posterior PDF by iteratively updating a set of random particles that are independently generated by multiple UAVs and distributed in the 3D searching space.
Each particle is characterized by its position and weight, where particle weight represents the probability of sampling this particle according to the PDF.
Moreover, the weighted summation of particles represents the
estimated source position.
We optimize the particle filter method to expedite particles convergence and improve localization efficiency considering the multi-UAV collaboration and real-time environmental states, where the updates of particles including particle weight, particle position, and the number of particles three parts, as described as follows.
Let $N_{i}^k$ denote the number of particles generated by UAV $i$ at the $k$-th iteration, and let $o_{i,n}^k$ and $\omega _{i,n}^k$ denote the particle position and the particle weight of the $n$-th particle generated by UAV $i$ at the $k$-th iteration.
Based on $o_{i,n}^k$ and $\omega _{i,n}^k$, the approximated posterior PDF of source position estimated by UAV $i$ can be expressed as follows:
\begin{equation}
p({r_{i,s}^k}|{d_i^{1:k}}) = \sum\nolimits_{n = 1}^{{N_{i}^k}} {{\omega _{i,n}^k}\delta (r_s - {o_{i,n}^k})},
\end{equation}
where $\delta$(.) represents a Dirac Delta function.
On this basis,
the particle weight, position and number will be updated iteratively to infer the next posterior PDF at the $k+1$-th iteration.


\subsubsection{\textbf{Particle Weight Update}}
The weight $\omega _{i,n}^k$ of the $n$-th particle is updated taking neighboring UAVs' confidence factors into account by UAV $i$ at the $k$-th iteration, which can be expressed as follows:
\begin{equation}
\widetilde \omega _{i,n}^k = \omega _{i,n}^{k - 1} \cdot p({d_i^k}|r_{i,s}^{k}) \cdot \prod\nolimits_{j = 1}^{{{\mathcal H}_i}} {{{[p(d_j^k|r_{i,s}^{k})]}^{\beta _{ij}^k}}}.
\label{eq:mw}
\end{equation}

Subsequently, the particle weight $\omega _{i,n}^k$ is normalized according to the following formula:
\begin{equation}
\omega _{i,n}^k = \frac{{\widetilde\omega _{i,n}^k}}{{\sum\nolimits_{n = 1}^{{N_{i}^k}} {\widetilde\omega _{i,n}^k}}}.
\end{equation}

\subsubsection{\textbf{Particle Position Update}}
Resampling alleviates the particle degradation, but it limits particle diversity and decreases the potential of exploration \cite{jouin2016particle}.
Here in addition to resampling,
we propose the Position Update combined with the Global and Local information (\textit{PUGL}) strategy to update particle position. 

At the beginning of the $k$-th iteration, the $N_{i}^k$ particles of UAV $i$ are firstly resampled according to the resampling condition.
Then, UAV $i$ randomly selects $N_{i,s}^k$ partial particles from $N_{i}^k$ particles to update their positions, while the remaining $N_{i}^k-N_{i,s}^k$ particles maintain unchanged in their original positions.
The $ N_{i,s}^k$ is computed based on preset maximum iteration number $k_{max}$ and the total particles number $N_{i}^k$, which can be given by as follows:
\begin{equation}
{N_{i,s}^k} = {N_{i}^k} \cdot {\left[ {1 - {{\left( {{{k}}/{{k_{max}}}} \right)}^{\gamma_1}}} \right]^{\gamma_2}},
\label{eq:select Np}
\end{equation}
where $\gamma_1$, $\gamma_2$ are coefficients.
When the iteration number $ k$ is small, the calculated $N_{i,s}^k$ is large, which means that more particles need to move for searching a wider area.
As $k$ increases, the calculated $N_{i,s}^k$ gradually decreases since more particles are closer to odor source.
After sampling $N_{i,s}^k$ particles, UAV $i$ firstly updates the positions of these particles under the environmental information sensed by itself (named as \textbf{\textit{env-pu}}), and then adjusts them in combination with multi-UAV collaboration (named as \textbf{\textit{col-pu}}).

\textbf{\textit{env-pu}}:
Inspired by the anemotaxis nature of living creatures when searching for food or sources, $N_{i,s}^k$ particles move upwind based on current positions and moving angles between particles' movement directions and the wind direction.
The position update of $o_{i,n}^k = [ x_{i,n}^k, y_{i,n}^k, z_{i,n}^k  ]$ at the $k$-th iteration is shown as follows:
\begin{equation}
    \begin{split}
    & x_{i,n}^k   = x_{i,n}^{k-1} + rand * x_{i,n}^{k-1}* cos(\theta_{i,n}^{k-1}), \\
    & y_{i,n}^k   = y_{i,n}^{k-1} + rand * y_{i,n}^{k-1}* sin(\theta_{i,n}^{k-1}), \\
    & z_{i,n}^k   = z_{i,n}^{k-1},
    \end{split}
    \label{eq:o1}
\end{equation}
where $rand$ represents a random value in [0,1], and $\theta_{i,n}^{k-1}$ is the moving angle between the $n$-th particle's movement direction of UAV $i$ and the current wind direction.
We assume that the wind direction is parallel to $x-y$ plane, and thus the value of $z_{i,n}^k$ maintains unchanged.

\textbf{\textit{col-pu}}: After being updated by \textit{env-pu} strategy, $N_{i,s}^k$ particles are further adjusted based on current cue-captured states.
We assume that a UAV is cue-captured, if it has encountered the emitted particulates by odor source during the period from the beginning to the present iteration.
Otherwise, this UAV is non-cue-captured, which means that no gas concentration information has been measured by this UAV up to now.


For the non-cue-captured UAV $j$, $N_{j,s}^k$ particles move towards the global optimal position $gbest_j^k$ with the random step sizes under the assistance of multi-UAV collaboration, which can be given by
\begin{equation}
   o_{j,n}^k \Leftarrow o_{j,n}^k + rand \cdot (gbest_j^k -  o_{j,n}^k),\forall n \in N_{j,s}^k.
   \label{eq:o3}
\end{equation}
$gbest_j^k$ denotes the weighted mean position of UAV $j$'s neighbors that are cue-captured, which can be expressed as
    $gbest_j^k =  {\sum\nolimits_{i = 1}^{\mathcal{H}_j^{cue}} {  \beta _{ji}^k  \cdot \mu _i^k   } }$,
where $\mathcal{H}_j^{cue}$ represents the UAV $j$'s neighboring UAVs set that are cue-captured, and $\mu _i^k$ represents the weighted mean position of $N_i^k$ particles for neighboring UAV $i$ at the $k$-th iteration.
$gbest_j^k$ plays an important role in guiding UAV $j$ to move towards the positions where it can encounter cues with a high possibility.

In the proposed \textit{PUGL} method, 
a better balance between exploration and exploitation can be achieved by dynamically adjusting the selected particles number in each iteration, and then the particle position update taking into account the environmental information and multi-UAV collaboration can accelerate particles convergence to the estimated odor source and improve the localization efficiency.

\subsubsection{\textbf{Particles Number Update}}
Traditional particle filter algorithms utilize a fixed number of particles to perceive environmental information and estimate source position.
Generally, utilizing a larger number of particles can improve the estimated accuracy of environmental states, but it has the drawback of high computation consumption.
Compared to a fixed number of particles, dynamically adjusting the number of particles based on the real-time searching states is a more efficient way to reduce computation overhead while maintaining localization efficiency.

We first determine the range of particles numbers based on the area size and source diffusion model.
The distance between two sampled particles should satisfy $g < 2\sqrt {D\tau }$ ($\sqrt {D\tau }$ is the \textit{typical length}  \cite{07infotaxis}), which guarantees that particle weights are updated in a reachable distance.
The minimum threshold of particles number $N_{\min}$ is calculated using $N_{\min} = {\frac{{Lx \cdot Ly \cdot Lz}}{{{g^3}}}}$.
The maximum threshold of particles number ${N_{\max}}$ is related to the convergence of search time, which is obtained by the numerical simulation in Sect.  \uppercase\expandafter{\romannumeral6}.

To dynamically update the particles number, cue-captured frequency and distance gap are taken into account.
We assume that the number of cues captured by UAV $i$ up to the $k$-th iteration is $m$.
Moreover, we define a constant number $\Delta c$, and define $k_{m-\Delta c}$ as the required iteration number of capturing $m-\Delta c$ cues, i.e., the $m-\Delta c$-th cue is captured at the $k_{m- \Delta c}$-th iteration.
We further define cue-captured frequency as the ratio of $\Delta c$ cues to the corresponding iteration difference.
$f_{i,cue}^k$ denotes the cue-captured frequency of UAV $i$ at the $k$-th iteration, which can be given by
    $f_{i,cue}^{k} =  {{\Delta c}}/{{k - k_{m-\Delta c}}}$.

$\sum\nolimits_{n = 1}^{{N_{i}^k}} {{\omega _{i,n}^k}{o_{i,n}^k}}$ represents the estimated source position using $N_i^k$ particles sampled by UAV $i$ at the $k$-th iteration.
Distance gap $f_{i,dist}^k$ is related to the distance between UAV $i$'s position and the estimated source position, which can be represented as
$f_{i,dist}^k = \exp \left( { -{1}/{{||u_i^k - {\sum\nolimits_{n = 1}^{{N_{i}^k}} {{\omega _{i,n}^k}{o_{i,n}^k}} } ||}}} \right)$.

As the number of iterations increases, UAV $i$ gradually approaches the odor source, and thus cue-captured frequency $f_{i,cue}^k$ increases and distance gap $f_{i,dist}^k $ decreases.
The number of particles for UAV $i$ at the $k$-th iteration can be represented as follows:
\begin{equation}
{N_{i}^k} = {N_{i}^{k-1}} * f_{i,dist}^k * (1 - f_{i,cue}^k).
\label{eq:Np}
\end{equation}

\subsection{Gaussian Fitting of Particles Distribution}

Multiple UAVs collaboratively find the odor source via sharing their local searching information.
We utilize 3D Gaussian distribution to fit particles distribution, and then the feature values of Gaussian distribution instead of burdensome particles information are exchanged among neighboring UAVs, significantly reducing the communication overhead.

In the initial phase, particles positions follow a random distribution and particles weights remain equal until the first gas measurement is obtained.
As more gas measurements are gradually detected, particles weights will be updated in each iteration to reflect the predicted source probability distribution, and thus weights distribution can be approximated by the 3D Gaussian distribution.
Then after the resampling and \textit{PUGL} method are implemented, the particles with larger weights are split into several new-generated particles, and the particles with lower weights are discarded, and thus the distribution of particles positions can also be regarded as a Gaussian distribution.

\begin{figure}[ht]
    \centering
    {\includegraphics[width=8cm]{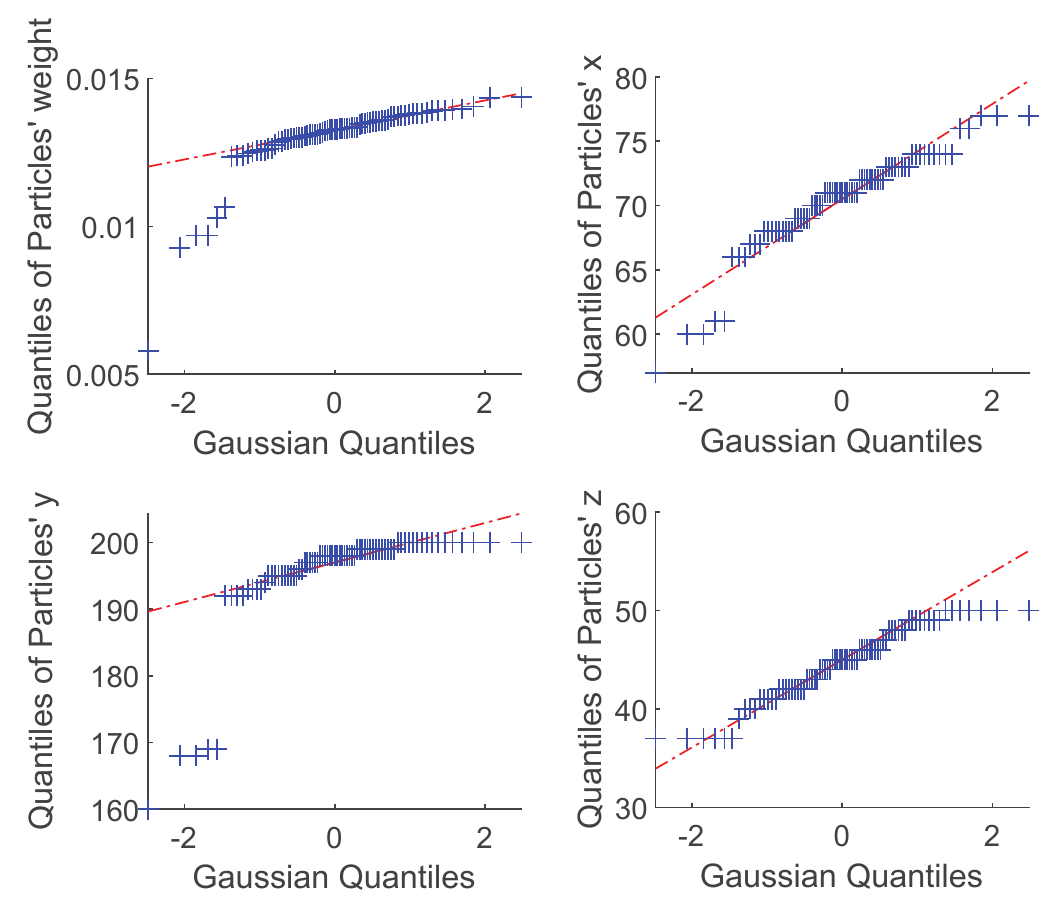}}
    \caption{A QQplot snapshot between Gaussian distribution and particles distribution.}
    \label{QQ plot}
\end{figure}

The Quantile-Quantile plot (QQplot) is a probability graphical method to measure the similarity between two probability distributions by comparing their quantiles \cite{qqplot}.
We use QQplot method to illustrate the similarity of particles weights distribution and particles positions distribution to Gaussian distribution respectively, as shown in Fig. \ref{QQ plot}.
It can be seen that the blue sampling points approximately lie on a straight line for the four subfigures, indicating that particle distribution, including positions distribution and weights distribution, is highly similar to the Gaussian distribution.

Therefore, the source probability distribution $ p(r_{i,s}^k)$ estimated by UAV $i$ at the $k$-th iteration can be approximated by the 3D Gaussian distribution, representing as $ p(r_{i,s}^k) \approx {\cal G}(\mu _i^k,\Sigma _i^k)$,
where $\cal G$ represents 3D Gaussian density function.
$\mu _i^k$ represents the mean value of Gaussian distribution, which is computed by the weighted average position of all particles sampled by UAV $i$ at the $k$-th iteration.
$\Sigma _i^k$ represents the covariance of Gaussian distribution, which denotes the dispersion degree of weighted particles at the $k$-th iteration.
The mean and covariance matrix are formalized as follows:
\begin{equation}
    \mu _i^k = {(\mu _{i,x}^k,\mu _{i,y}^k,\mu _{i,z}^k)^T},
\end{equation}
\begin{equation}
\Sigma _i^k = \left[ {\begin{array}{*{20}{c}}
{{{(\sigma _{i,x}^k)}^2}}&{\rho \sigma _{i,x}^k\sigma _{i,y}^k}&{\rho \sigma _{i,x}^k\sigma _{i,z}^k}\\
{\rho \sigma _{i,x}^k\sigma _{i,y}^k}&{{{(\sigma _{i,y}^k)}^2}}&{\rho \sigma _{i,y}^k\sigma _{i,z}^k}\\
{\rho \sigma _{i,x}^k\sigma _{i,z}^k}&{\rho \sigma _{i,y}^k\sigma _{i,z}^k}&{{{(\sigma _{i,z}^k)}^2}}
\end{array}} \right].
\end{equation}

After the above Gaussian fitting process, the values of ${\mu _i^k}$ and $ {\Sigma _i^k}$ computed by UAV $i$ are exchanged with its neighbors, greatly reducing communication burden compared with transmitting particle positions and weights.
Furthermore, based on $[{\mu _i^k,\Sigma _i^k}]$ of UAV $i$ and received $[{\mu _j^k,\Sigma _j^k}]$ from neighboring UAV $j$, the computation of KL divergence and confidence factor $\beta_{ij}^k$  is also be simplified.


\subsection{The Flow of Collaborative Particle Filter Algorithm}

Basic running steps for the collaborative particle filter algorithm of UAV $i$ at the $k$-th iteration in the source estimation phase are summarized in Algorithm 1.
The inference of the source probability is driven by iteratively updating the particles.
For each particle, the weight is updated based on the importance sampling considering the multi-UAV collaboration to represent the predicted source probability distribution  in line 1$\sim$2.
Then, the resampling is implemented to deal with the particle degradation in line 4.
Then the particle position is updated using \textit{PUGL} strategy for accelerating the particle convergence to the estimated source position, as described in lines 6$\sim$7.
Next, the number of particles is updated adaptively to reduce the computation burden in line 8.
Moreover, the fitting values of particles distribution are obtained in line 9 to compute the confidence factor.
Finally, the outputs of collaborative particle filter include the updated particles positions, weights, number and fitting values to start the next iteration.

\begin{algorithm}
\caption{Collaborative Particle Filter Algorithm for UAV $i$ at the $k$-th iteration}
\LinesNumbered
\KwIn{current measurement, particles positions, weights, and particles number at the $k$-th iteration of UAV $i$}
    \ForEach{$n \in N_{i}^k$ }{
        update particle weight: $ w_{i,n}^{k+1} \leftarrow w_{i,n}^{k}$ using (\ref{eq:mw})\;
    }
    normalize the particles weights\;
    resample particles using the low variance sampling\;
    compute the selected particles number $N_{i,s}^k$ using (\ref{eq:select Np})\;
    \ForEach{$n \in N_{i,s}^k$ }{
        update particle position via \textit{PUGL}: $o_{i,n}^{k+1} \leftarrow o_{i,n}^{k}$ using (\ref{eq:o1})(\ref{eq:o3})\;
    }
    update particles number: $ N_{i}^{k+1} \leftarrow N_{i}^{k}$  using (\ref{eq:Np})\;
    obtain the  $[{\mu _i^k,\Sigma _i^k}]$ by fitting the particles distribution\;
\KwOut{$[{\mu _i^k,\Sigma _i^k}]$, the particles positions, weights, and particles number at the iteration $k+1$ of UAV $i$}
\end{algorithm}

\section{Navigation: POMDP-based Multi-UAV Adaptive Path Planning}
In this section, we mainly focus on the navigation phase in the probability-based odor source localization framework.
After obtaining the inferred source probability map under the multi-UAV collaboration, each UAV determines its next moving position independently with the objective of minimizing the overall search time.
The path planning of each UAV is a sequential decision making process, which includes the decision making for flying direction and moving step.
Both of the decisions can be described as POMDP, since each UAV has partial knowledge of the searching states and environment perception.
POMDP mainly includes three elements: the information state, the set of admissible actions, and the reward function.
Each action is mapped into a real-number by the reward function based on current information state.

\subsection{Decision Making for Flying Direction}
Each UAV determines its flying direction based on POMDP, and
the information state at each iteration is represented by the current posterior PDF of source parameters inferred by each UAV.
The action set of each UAV is constructed by flying directions.
Assuming that the whole searching space is gridded, the number of each UAV's flying directions is
the neighboring grid direction of the current position in 3D space, i.e., 26 directions.
Considering that the uncertainty reduction in source probability inference and the distance changing between the UAV and estimated source position are highly different when the UAV flies in different directions,
we define the reward function by jointly considering entropy reduction and distance variation to determine the flying direction in the current path planning. 

Specifically, the entropy of UAV $i$ at the $k$-th iteration is computed as follows:
\begin{equation}
    H_i^k = - \sum\nolimits_{n = 1}^{{N_i^k}} {\omega _{i,n}^k\log \omega _{i,n}^k}.
\end{equation}
The entropy reduction $\Delta H = H_i^k -H_i^{k+1}$ between the $k$-th iteration and the $k+1$-th iteration is used to measure the degree of uncertainty reduction between adjacent iterations.
The spatial distance between UAV $i$ and the estimated source position at the $k$-th iteration is represented as follows:
\begin{equation}
D_i^k = ||u_i^k - \sum\nolimits_{n = 1}^{{N_i^k}} {\omega _{i,n}^k o_{i,n}^k} ||,
\end{equation}
where $||.||$ is L2 normalization.
As UAV $i$ moves closer to the estimated source, the distance $D_i^k$ constantly decreases.

We further define the value function $W_i^k$ of UAV $i$ at the $k$-th iteration as a weighted sum of spatial distance and information entropy, which is expressed as follows:
\begin{equation}
\begin{aligned}
    W_i^k &= D_i^k +  h_1 * H_i^k  \\
    &= ||u_i^k - \sum\nolimits_{n = 1}^{{N_i^k}} {\omega _{i,n}^k o_{i,n}^k} || - h_1 * \sum\nolimits_{n = 1}^{{N_i^k}} {\omega _{i,n}^k\log \omega _{i,n}^k},
\end{aligned}
\label{eq:dir1}
\end{equation}
where $h_1$ denotes the weight of information entropy, and it is positively correlated with the variance of particle distribution.
As particles gradually converge to the source position, the uncertainty of source probability reduces and $h_1$ decreases, which means that UAV $i$ will move towards the estimated source position under the guidance of spatial distance to a large extent.

After obtaining corresponding values of different directions in the action set, the reward function is defined as the difference of value function between the $k$-th and the $k+1$-th iterations.
Since the measurement $d_i^{k+1}$ at the $k+1$-th iteration is unknown, we use the approximated expectation $E[W_i^{k+1}]$ for UAV $i$ to represent the value function at the $k+1$-th iteration, which can be calculated by:
\begin{equation}
    E[W_i^{k+1}] = \sum\nolimits_{d=0}^{d_{max}} {p(d_i^{k+1}) W_i^{k+1} },
\end{equation}
where $d_{max}$ represents the maximum measurement following the Poisson distribution.
$d_{max}$ can be determined when $p(d_{max})$ highly approaches 1.
Then, the action $ v_{i,dir}^k$ that UAV $i$ chooses at the $k$-th iteration is the direction with the maximum reward, which can be formulated as
\begin{equation}
    v_{i,dir}^k = \arg\max\limits_{v_{i,dir}^k \in \mathcal{V}_{dir}} \{W_{i}^k  - E[W_i^{k+1}]\}.
    \label{eq:dir2}
\end{equation}

\subsection{Decision Making for Moving Step}

After the flying direction is chosen by each UAV, the moving step is then determined as below.
If adopting a fixed step in the navigation phase, a large step allows the UAV to move fast, but easily results in path redundancy;
while a small step makes the UAV move slowly and consequently prolongs the search time, which is suitable when the UAV approaches the odor source.
Therefore, it is necessary to dynamically adjust the moving step at each iteration based on real-time states rather than maintaining a fixed step throughout the whole localization procedure, aiming to reduce the overall search time while maintaining the localization accuracy.
Moreover, adaptive adjustment of moving step can also reduce the number of turning points and hovering points of the UAV compared with the fixed step, and thus efficiently reducing movement energy consumption.

Each UAV determines its moving step based on POMDP,
and the information state at the $k$-th iteration includes the posterior PDF of source parameters and the flying direction obtained by the above part.
The action set of each UAV is constructed by its moving steps, which is denoted by $\mathcal{V}_{step}=[1, \cdots, l_{max}]$.
The minimum step is set as 1, and the maximum step $l_{max}$ is determined by the searching area size and the source diffusion model.
To determine the value of $l_{max}$,
here we define the metric $\zeta$ as the ratio of source diffusion area size $A_{source}$ to the entire searching area size $A_{whole}$, which can be represented as follows:
\begin{equation}
 \zeta=\frac{{A_{source}}}{{A_{whole}}}=\frac{G_{c>th}}{Lx*Ly*Lz},
\end{equation}
where $A_{source}$ is approximated by the number of grids whose concentration is larger than the preset concentration threshold $ G_{c>th}$ in the whole grid area.
It can be seen that $\zeta < 1$.
A smaller $\zeta$ is obtained when the searching area is larger or the odor source diffusion area is relatively smaller, and then a larger maximum step $l_{max}$ will be adopted to explore this area.
We define that
if $\zeta < 0.1$, $l_{max}$ is calculated depending on $\zeta*l_{max}=0.1$.
Otherwise, $l_{max}$ is fixed at 1, which corresponds to the case of fixed step.

The reward to determine the moving step size is related to the historical measurement information of the UAV.
Let $u_i^{k+1}(l)$ denote the next 3D position of UAV $i$ given moving step $l$,
and let $ c_i^k(l)$ denote the sphere whose diameter is the difference value between the current position $u_i^k$ and the next position $u_i^{k+1}(l)$ of UAV $i$ and both points are on the sphere.
A larger step corresponds to the sphere with a larger diameter.
We further denote $z_i^k(l)$ to represent the number of historical measurement points contained in sphere $c_i^k(l)$, which reflects the exploration ability of UAV $i$ with moving step $l$.
Given two adjacent moving steps $l_2$ and $l_1$, we assume that $l_2>l_1$, and define $Z_i^k(l_2) = z_i^k(l_2)-z_i^k(l_1)$ as the difference in the number of historical measurement points between sphere $c_i^k(l_2)$ and sphere $c_i^k(l_1)$, which represents the historical measurement points that are contained only by the larger sphere $z_i^k(l_2)$ but not contained by the smaller sphere $z_i^k(l_1)$.
If $l=1$, the number of measurement points can be represented as $ Z_i^k(l) = z_i^k(l)$.
A moving step with smaller $Z_i^k(l)$ means that the distance between $u_i^{k+1}(l)$ and the positions that UAV $i$ has already flown is large, and thus UAV $i$ tends to explore the unknown area.

We design the reward function based on the number of historical measurement points and spatial distance between each UAV and the estimated source position to determine the moving step of each UAV in current path planning.
The reward obtained by UAV $i$ moving $l$ step at the $k$-th iteration can be represented as follows:
\begin{equation}
   R_i^k(l) = D_i^k(l) + h_2 * Z_i^k(l),
   \label{eq:step1}
\end{equation}
where $h_2$ is the weight of historical measurement points at the $k$-th iteration.
As UAV $i$ gets close to the estimated source, $h_2$ is decreased to weaken the effect of $Z_i^k(l)$, and UAV $i$ tends to reduce the exploration opportunity.
The action of UAV $i$ at the $k$-th iteration is chosen with the minimum reward value, which is represented as follows:
\begin{equation}
    v_{i,step}^k = \arg \min\limits_{{v_{i,step}^k} \in \mathcal{V}_{step} } R_{i}^k(l).
    \label{eq:step2}
\end{equation}

\subsection{The Flow of POMDP-Based Multi-UAV Path Planning}

Each UAV individually plans its path including flying direction and moving step at each iteration.
Algorithm 2 describes the whole path planning process of UAV $i$ at the $k$-th iteration in the navigation phase.
In each iteration, the flying direction of UAV $i$ is firstly chosen, and then the moving step in this direction of UAV $i$ is determined.
Both two decision making processes can be modelled as POMDP.
Specifically, the reward function related to flying direction is designed based on the entropy reduction and distance changing as described in lines 4$\sim$7.
Then the UAV $i$ chooses its direction with the maximum reward to approach the odor source in line 8.
After the flying direction is determined, the measurement points and distance information are considered to construct the reward function in regard to moving step as described in lines 10 $\sim$ 12.
Then the UAV $i$ chooses its moving step with the minimum reward in line 13.
The output of this algorithm is the next position of UAV $i$ at the $k+1$-th iteration.

\begin{algorithm}
\caption{Adaptive path planning for UAV $i$ at the $k$-th iteration}
\LinesNumbered
\KwIn{ UAV $i$' position, source PDF, $\zeta$}
Create the action set of admissible flying directions $ \mathcal{V}_{dir}$\;
Create the action set of admissible moving steps $ \mathcal{V}_{step}$\;
\ForEach{$v_{i,dir}^k \in \mathcal{V}_{dir}$ }{
    compute the information entropy $H_i^k$\;
    compute the distance $D_i^k$ between UAV $i$ and estimated source position\;
    compute the value function $W_i^k$ using (\ref{eq:dir1})\;
    compute the reward function $ W_i^k-E [W^{k+1}_i]$\;
}
choose the flying direction with the maximum reward using (\ref{eq:dir2})\;
\ForEach{$v_{i,step}^k \in \mathcal{V}_{step}$ }{
    compute the number of measurement points $Z_i^k$\;
    compute the distance $D_i^k$ between $u_i^ {k+1}$ and the estimated source position\;
    compute the reward function $ R_i^k$ using (\ref{eq:step1})\;
}
choose the moving step with the minimum reward using (\ref{eq:step2})\;
\KwOut{the next position of UAV $i$ at the $k+1$-th iteration }
\end{algorithm}

Algorithm 3 describes the algorithm of the whole odor source localization for each UAV with the collaboration of neighbors, and uses UAV $i$ as an example.
UAV $i$ communicates with its neighboring UAVs to measure the cognitive difference of inferred source parameters as described in lines 5$\sim$7.
On this basis, Algorithm 1 is called to estimate the real-time source probability map.
After obtaining source inferences, Algorithm 2 is called to determine the next position of UAV $i$.
Next, UAV $i$ flies to next position to detect new measurement and start the next iteration until the source declaration condition is satisfied.
This algorithm is exited with the outputs including the estimated source position and the flying trajectories of multiple UAVs.

\begin{algorithm}[ht]
\caption{Odor Source Localization Algorithm for UAV $i$}
\LinesNumbered
\KwIn{the area size, the measurement model, the communication radius}
randomly distribute the particles in the searching area\;
deploy UAV $i$ in the corner of searching area\;
\While{ $k \le k_{max}$  and source declaration condition not satisfied }{
    obtain measurement $d_i^k$ at the current position $u_i^k$\;
    \ForEach{UAV $j \in \mathcal{H}_i$ }{
        share local information [$\mu _i^k,\Sigma _i^k$] to UAV $j$\;
        receive the information from UAV $j$, and then compute confidence factor $\beta_{ij}^k$\;
    }
    update particles information and estimate the source probability based on $\beta_{ij}^k$ via Algorithm 1\;
    determine the next position of UAV $i$ via Algorithm 2\;
    move to the new position $u_i^{k+1}$\;
    $k = k+1$\;
}
\KwOut{the odor source coordinates and the flying trajectory of UAV $i$}
\end{algorithm}

\section{Performance Evaluation}
In this section, two plume simulation platforms are considered to assess the performance of the proposed \textit{MUC-OSL} algorithm via extensive Monte Carlo simulations.
The first simulation is conducted using rate-based plume model
in a simplified simulator.
Then, we implement the algorithm in the high-fidelity robotic simulator, GADEN \cite{monroy2017gaden}, to validity its effectiveness in a more realistic scenario.
It is worth noting that the propellers on UAV will greatly impact the plume distribution around the UAV
\cite{ercolani20203d},
which is not considered in this paper as most previous works focused on the localization algorithm
\cite{rossi2014gas, kersnovski2017uav, duisterhof2021sniffy, access19multiUAV}.

\subsection{Simulation Setting}

The algorithm and environment parameters for both simulation platforms are listed in Table \ref{tab1}.
The initial positions of all UAVs are at the lower left corner of the searching area as in previous work \cite{20-IJRR}.
We assume that all UAVs are homogeneous and have equivalent resources and battery capacities, and the maximum search time of each UAV is 20 minutes considering the limited energy of UAVs.

\begin{table}[]
\small
\caption{Parameters setting for the algorithm and plume model.}	
\label{tab1}
\resizebox{\linewidth}{!}
{
\begin{tabular}{l | l p{4cm} l}
\hline
& \textbf{Symbol} & \textbf{Description}                            & \textbf{Value} \\ \hline
\multirow{7}{*}{\begin{tabular}[c]{@{}l@{}}\\Algo\\param\end{tabular}}
& $\gamma_1$             & Parameter of $N_{i,s}$ calculation                  & 1.8       \\
& $\gamma_2$             & Parameter of $N_{i,s}$ calculation                  & 4         \\
& $H$                  & The number of UAVs                              & 3               \\
& $N$                  & Default particles number                        & 100             \\
& $N_{max}$              & Maximum particles number                        & 160            \\
& $N_{min}$              & Minimum particles number                        & 20             \\
& $k_{max}$              & Maximum iterations number                        & 800            \\
\hline

\multirow{4}{*}{\begin{tabular}[c]{@{}l@{}}Rate-\\based\\plume\\param\end{tabular}}
& $Q$                  & Release rate of diffusion model                 & 5 kmol/(m\textsuperscript{2}s)              \\
& $V$                  & Mean wind velocity                              & 1 m/s          \\
& $D $                 & Diffusion coefficient                           & 1              \\
& $\tau$                & Particle lifetime                               & 100            \\ \hline

\multirow{8}{*}{\begin{tabular}[c]{@{}l@{}}\\Gaden-\\based\\plume\\param\end{tabular}}
& $T_p$                  & Temperature                                     & 298 K            \\
& $P$                    & Pressure                                        & 1 Atm             \\
& $\Delta T_g$           & Time step for plume updates                     & 0.5 s       \\
& $F_s$                  & Number of filaments released each second        & 50 filaments/s              \\
& $F_c$                  & Gas concentration at the center of the filament & 100 ppm            \\
& $F_\sigma$             & Range of the initial filament                   & 50 cm            \\
& $\gamma_{g}$           & Growth ratio of the $F_\sigma$                  & 50 cm\textsuperscript{2}/s         \\
& $F_R$                  & Range of the white noise added on each iteration   & 0.5 m            \\ \hline
\end{tabular}
}
\end{table}

\noindent \textbf{Baselines.}
To verify the performance of the proposed \textit{MUC-OSL} algorithm,
three baselines are considered by replacing partial component of \textit{MUC-OSL}.
\begin{itemize}
\item \textit{Col-Inf}: It uses the traditional particle filter for estimation, and plans UAVs' paths based on POMDP with fixed moving step for navigation \cite{RAS2020song}.
    It is a collaborative infotaxis algorithm and we renamed it as \textit{Col-Inf}.

\item \textit{Col-PF}: It uses the proposed \textit{Col-PF} strategy for  estimation, and plans UAVs' paths based on POMDP with fixed moving step for navigation.

\item \textit{Adap-PP}: It uses the traditional particle filter for estimation and the proposed \textit{Adap-PP} strategy for navigation.

\end{itemize}

\noindent \textbf{Evaluation Metrics.}
The mean search time (MST) and the success rate (SR) are considered as evaluation metrics to verify the effectiveness of the proposed algorithm.
The search time represents the time spent by the first UAV reaching the odor source in a simulation.
The mean search time denotes the average value for the search times of 200 Monte Carlo simulations.
It should be emphasized that the search time cannot be expressed by the number of iterations like previous works using a fixed moving step \cite{Inffus16study,RAS2020song}, since the variable moving step enables the search time to be not positively related with the number of iterations.
Specifically, considering the kinetics features of UAVs, we define the search time of a UAV as the summation of flying time, hovering time, and turning time in a more practical view.
In our simulation, we assume that the flying speed of UAVs is fixed as 1m/s, the hovering time of UAVs for each hovering point is fixed as $1s$, and the turning time of UAVs for each turning point is $1s$.


Moreover, the success rate is defined as the ratio of the number of successful source localizations to the number of total Monte Carlo simulations, and the odor source is successfully found only if the source declaration condition is satisfied within the maximum iteration threshold.

\subsection{Results for simplified simulator}

\begin{figure*}[t]
	\begin{center}
		\subfigure[running with k=0]{
			\includegraphics[width=4cm]{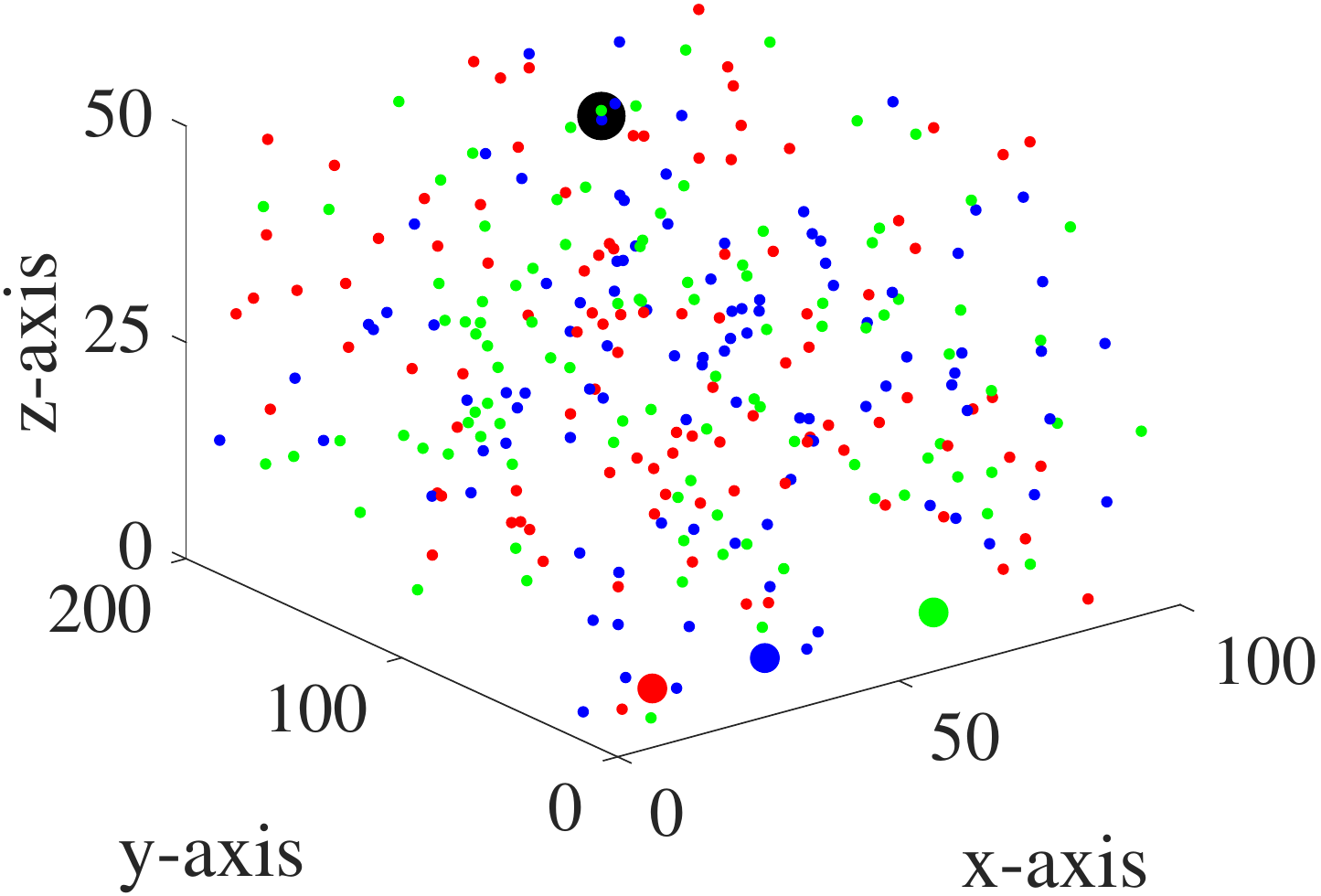}
		}
        \subfigure[running with k=46]{
			\includegraphics[width=4cm]{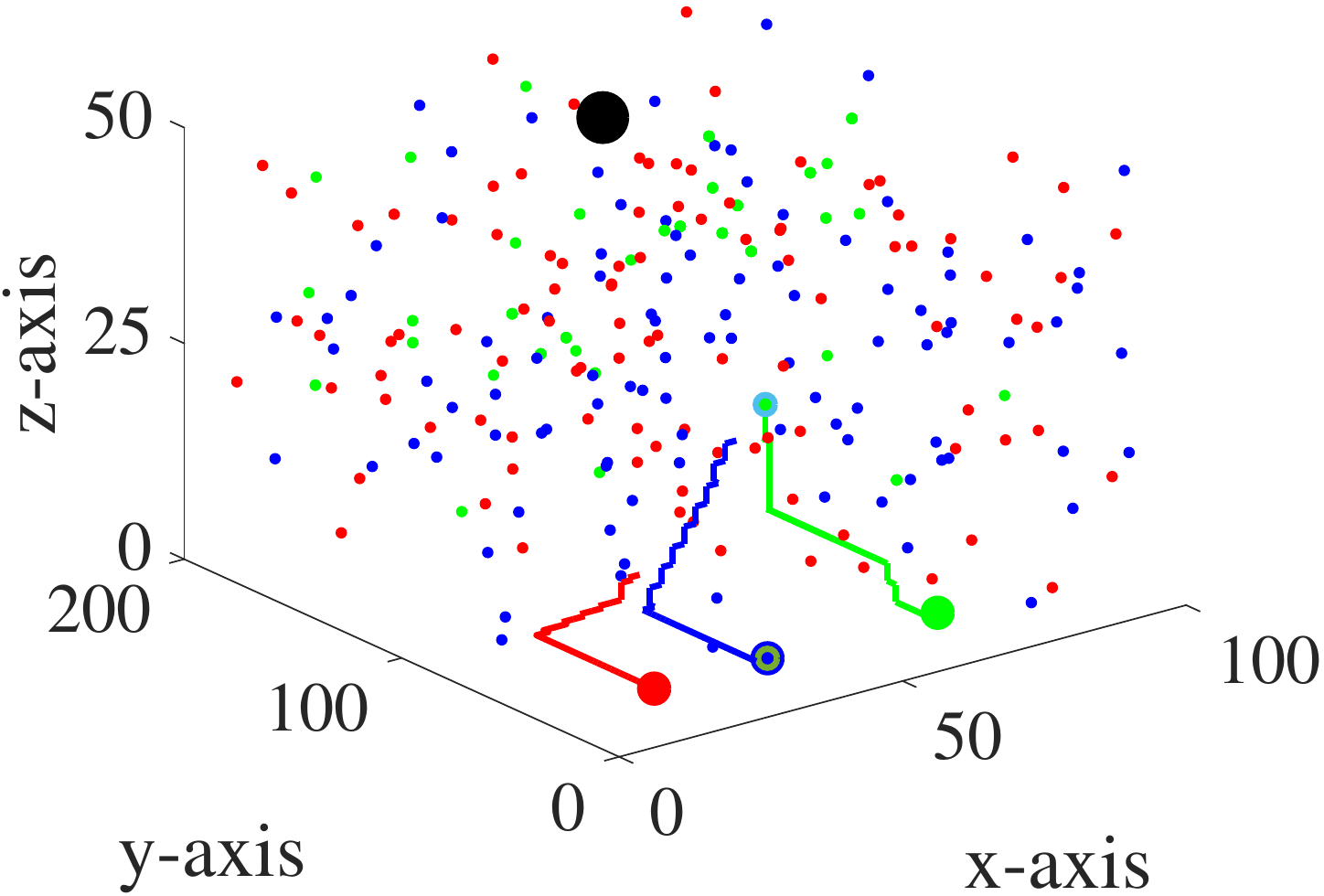}
		}
		\subfigure[running with k=102]{
			\includegraphics[width=4cm]{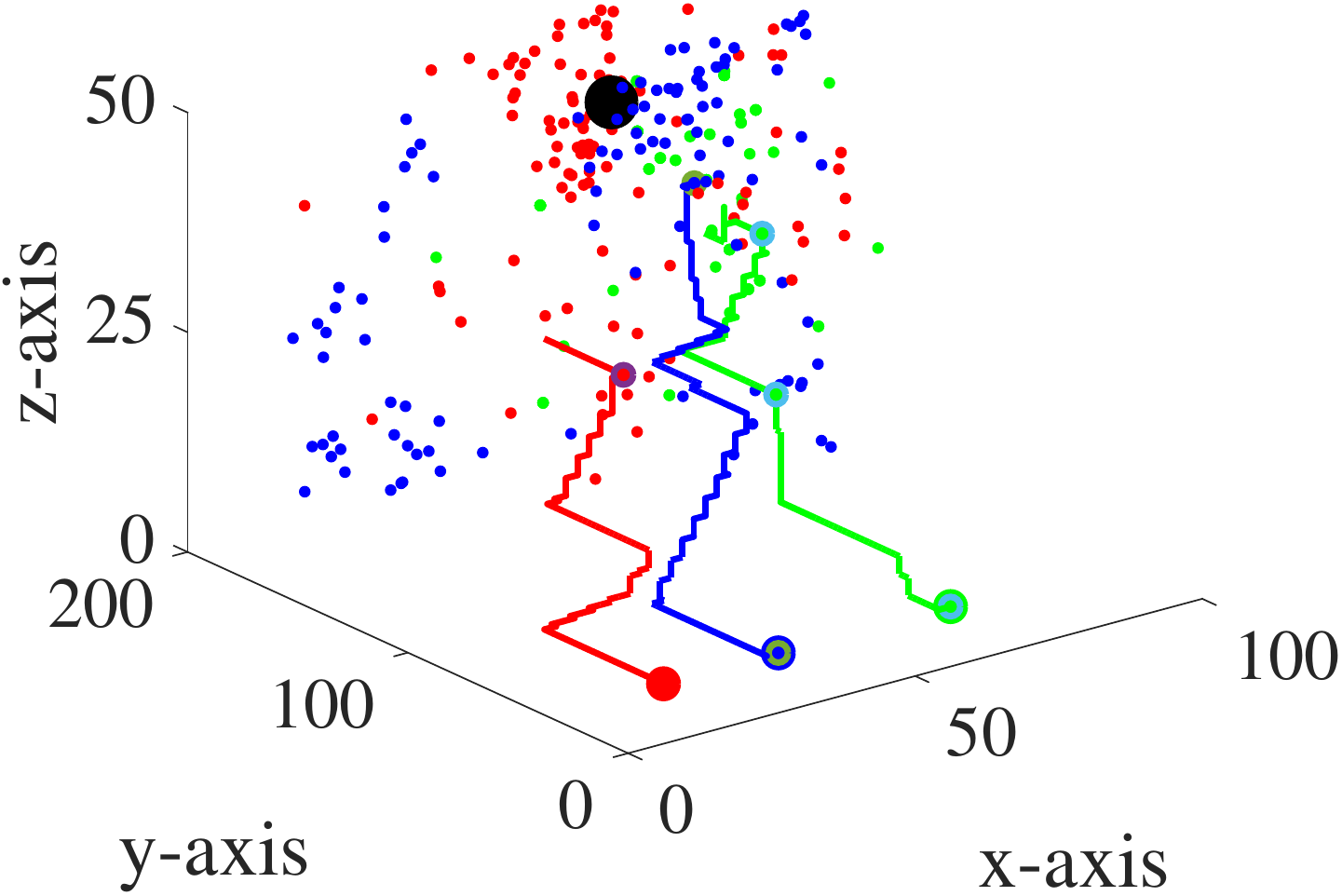}
		}
		\subfigure[running with k=168]{
			\includegraphics[width=4cm]{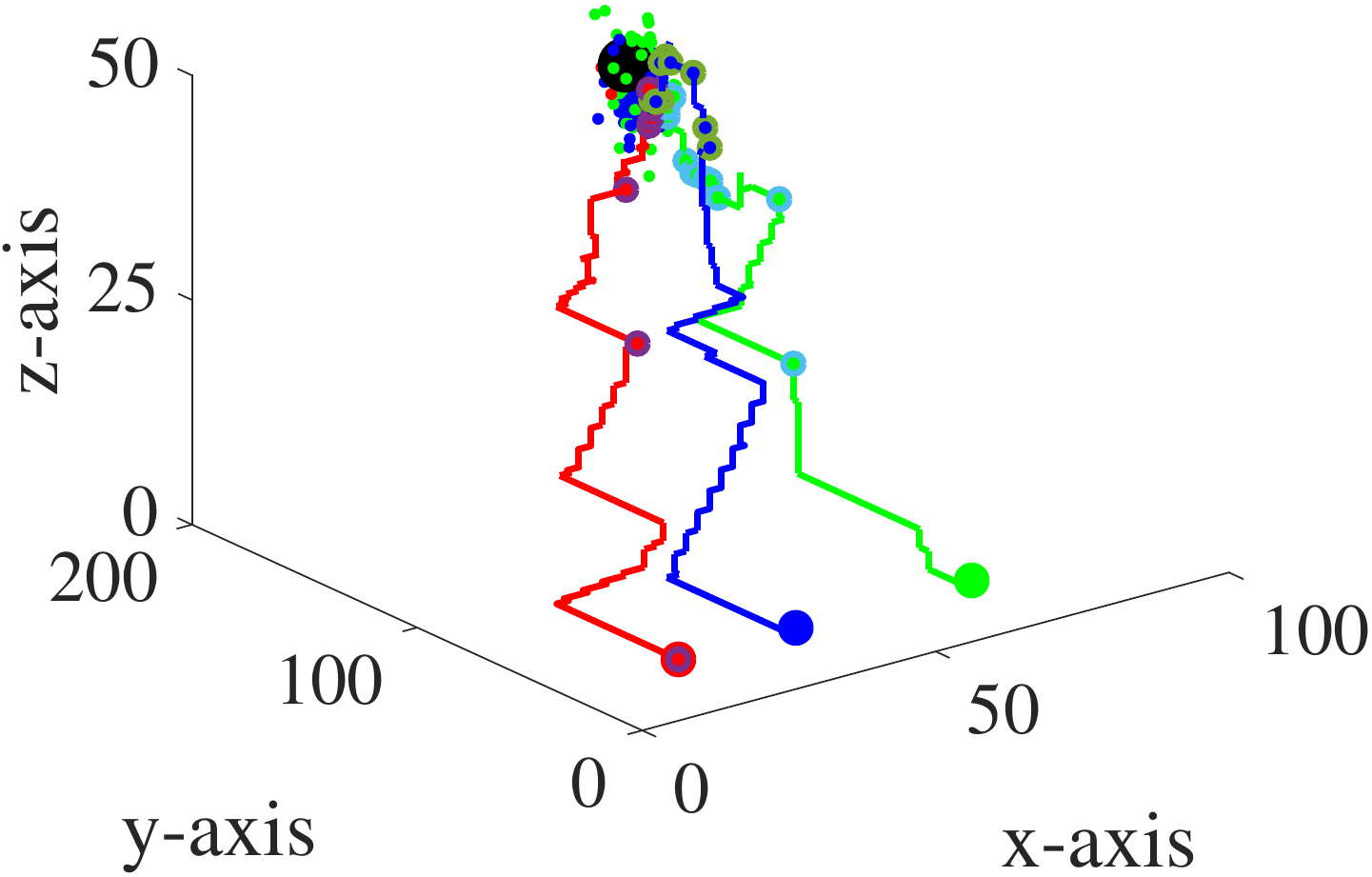}
		}
		\caption{A running process example for 3D space}
		\label{running-3d}
	\end{center}
\end{figure*}

The illustrative running results of odor source localization  under three UAVs collaboration using the proposed algorithm in 100m*200m*50m searching space are represented in Fig. \ref{running-3d}.
Fig. \ref{running-3d}(a) shows the initial states when $k=0$, where black dot in the upper right corner indicates the position of odor source. Initially, three UAVs are deployed in the bottom left corner represented using different colored dots, which are [10m,10m,5m] (UAV 1 in red), [30m,10m,5m] (UAV 2 in blue), and [60m,10m,5m] (UAV 3 in green). The particles represented by the small dots are randomly distributed in the whole area in the inception, and the particles with different colors belong to different UAVs. Then three UAVs move to find the plume based on the sensor measurements and collaboration information.

Fig. \ref{running-3d}(b) represents that UAV 3 first captures the non-zero measurement among all UAVs, and the particles of UAV 3 are resampled at the $46$-th iteration.
Then, all UAVs have captured cues when $k=102$, and all particles move closer to the estimated source position under resampling and \textit{PUGL} strategies, as shown in Fig. \ref{running-3d}(c). Fig. \ref{running-3d}(d) represents the searching paths obtained after all UAVs arriving the odor source when $k=168$ and all particles have also converged to the source position. Three lines in different colors represent the flight paths of three UAVs, and the dots on the lines indicate that a cue has been captured at this position. It can be seen that more cues will be captured near the odor source.

In the following part, we compare our proposed algorithm with three baselines in terms of MST and SR.
Moreover, we also evaluate the impact of some important parameters on the performance of source localization, such as the size of source searching area, the number of UAVs, and the number of particles, etc.

\begin{figure}[ht]
	\begin{center}
		\subfigure[mean search time]{
			\includegraphics[width=4cm]{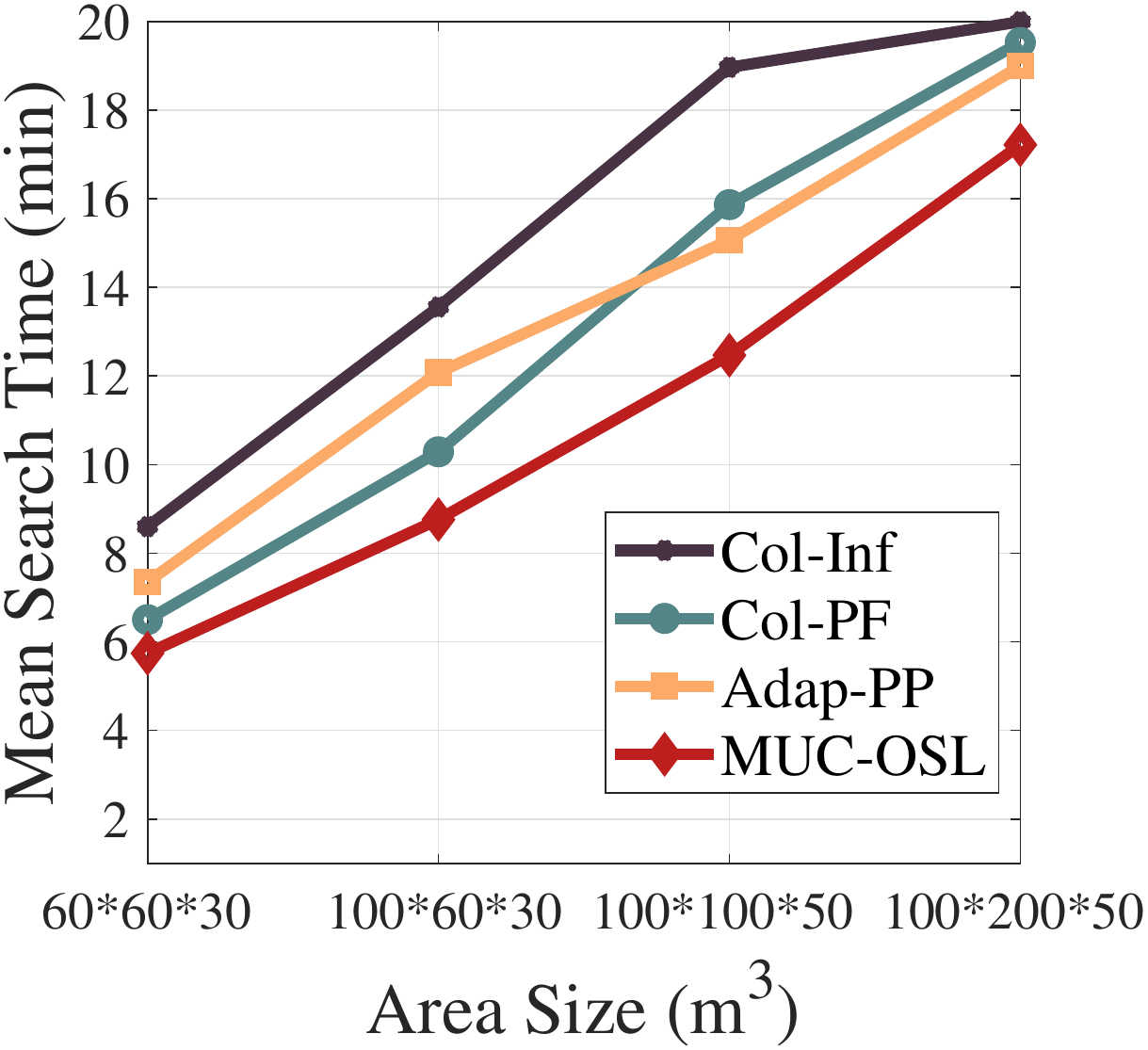}
		}
        \subfigure[success rate]{
			\includegraphics[width=4cm]{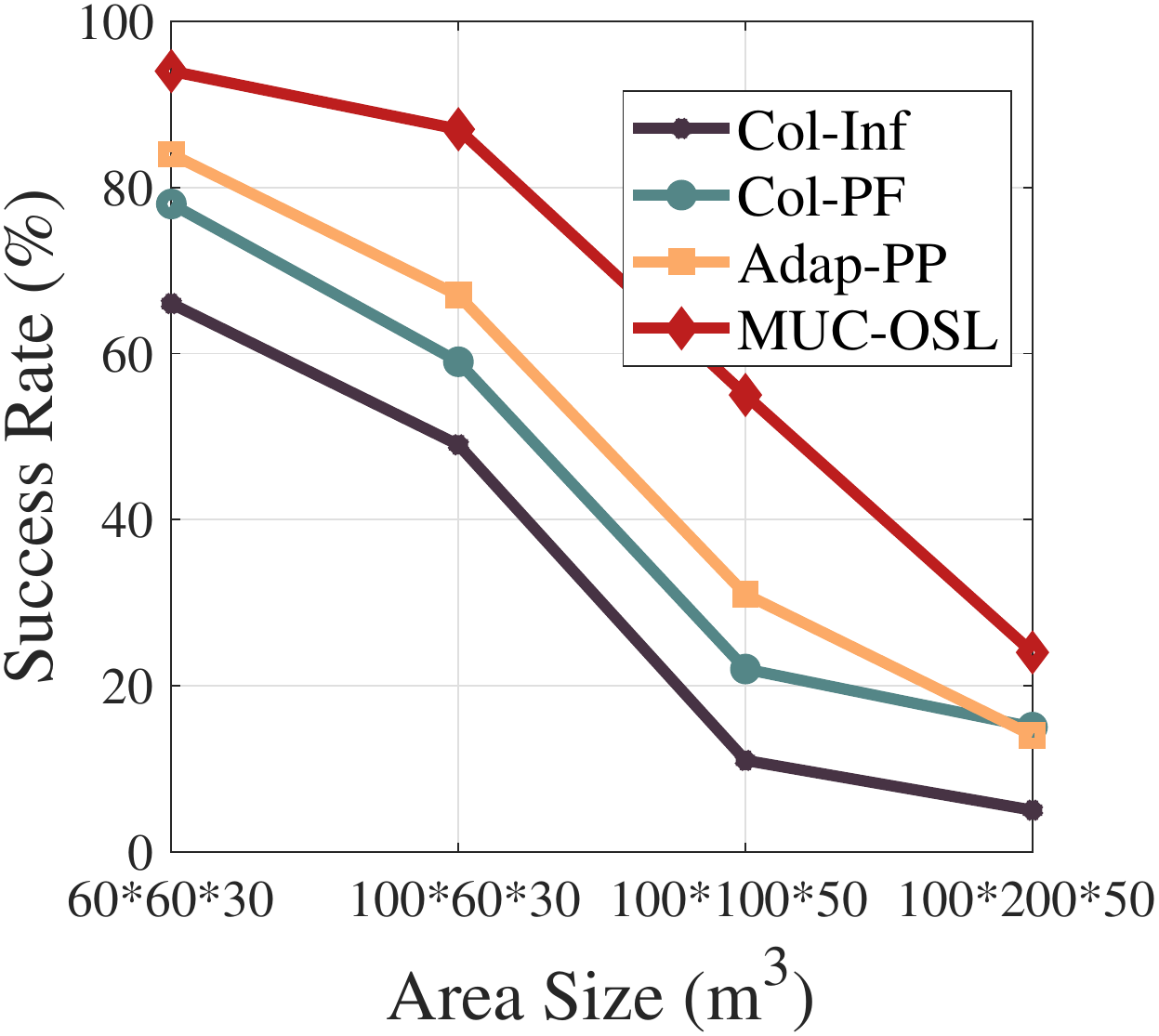}
		}
		\caption{Performance comparison of four algorithms with different searching area sizes.}
		\label{resultArea}
	\end{center}
\end{figure}

\textbf{\textit{1) Different searching area sizes:}}
We study the performance of four algorithms as the size of searching area varies as shown in Fig. \ref{resultArea}.
The simulations are conducted for searching areas: 60m*60m*30m, 100m*60m*30m, 100m*100m*50m and 100m*200m*50m.
In each case, 200 Monte Carlo simulations are conducted to obtain the MST and the SR.
Three UAVs are used and placed in the lower left corner of the searching area.
As expected, as the searching area decreases, the mean search time decreases and the success rate improves for all algorithms.
\textit{MUC-OSL} has a better success rate and shorter search time for all the searching areas compared with the baselines.
The required search time and success rate of \textit{Col-Inf} vary greatly as the searching area size increases, which indicates \textit{Col-Inf} is more sensitive to the changes of region size.
The \textit{Adap-PP} takes less time than \textit{Col-PF} when the searching area size increases from 100m*60m*30m to 100m*100m*50m, which verifies that adaptive determination of UAVs' moving directions and step sizes will significantly reduce the search time when the searching area is larger.
In the searching area of 100m*200m*50m, \textit{MUC-OSL} almost exhausted the energy achieving only 24\% success rate, while \textit{Col-Inf} could not accomplish localization task under the energy constraint of the UAV.

\begin{figure}[ht]
	\begin{center}
		\subfigure[mean search time]{
			\includegraphics[width=4cm]{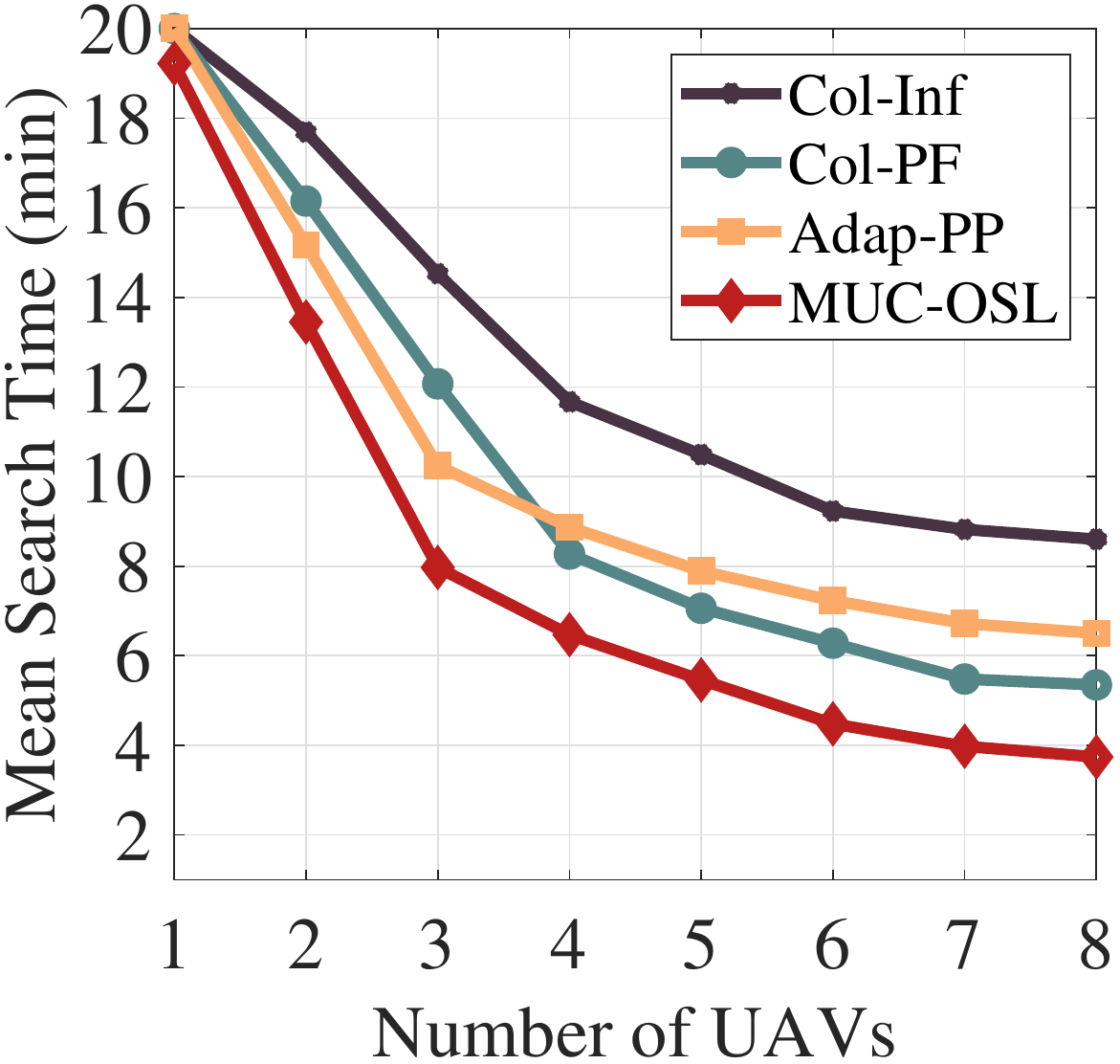}
		}
        \subfigure[success rate]{
			\includegraphics[width=4cm]{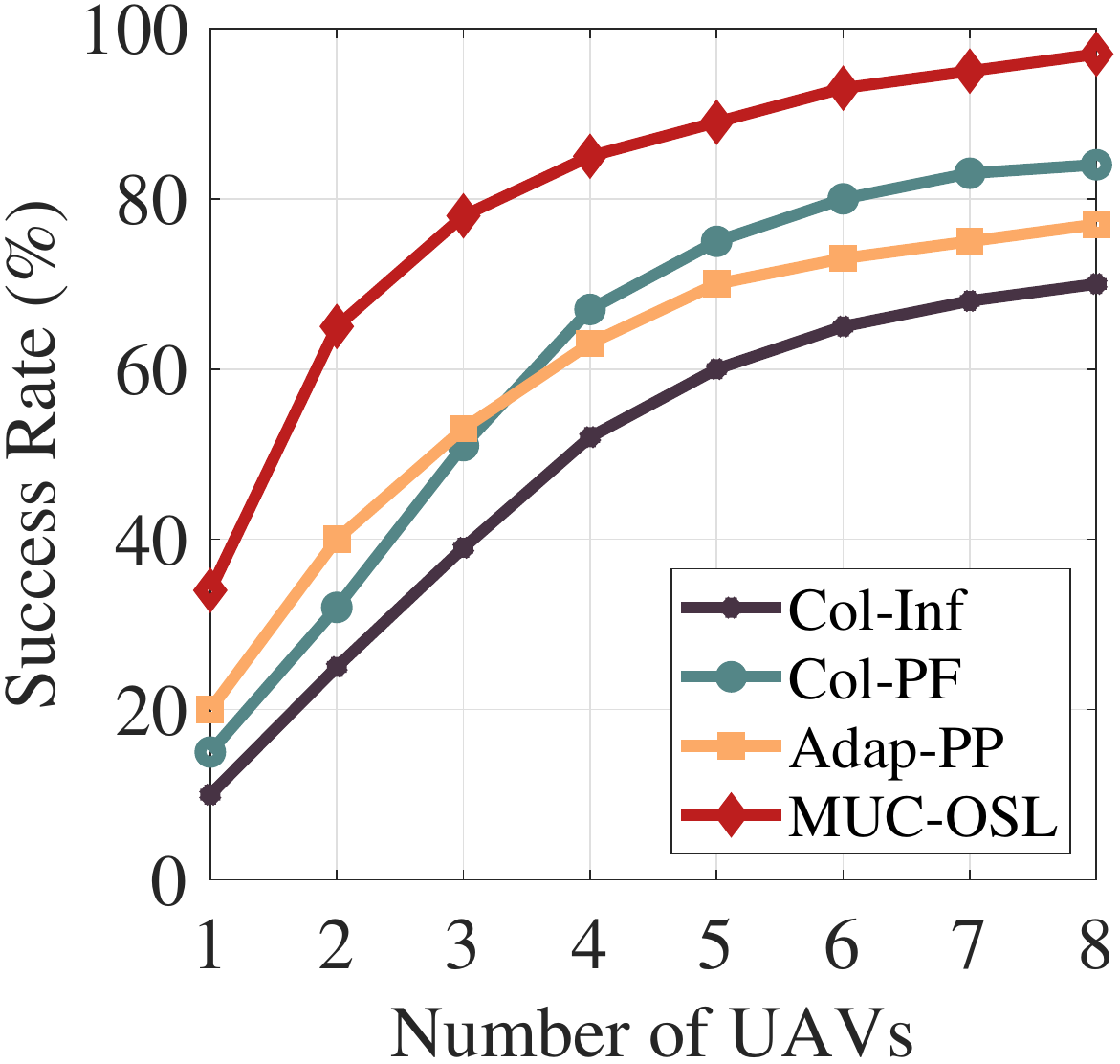}
		}
		\caption{Performance comparison of four algorithms with different numbers of UAVs.}
		\label{resultUAV}
	\end{center}
\end{figure}

\textbf{\textit{2) Different UAVs numbers:}}
We observe the impact of the number of UAVs on the MST and SR in this simulation, and the size of UAV teams varies from 1 to 8. 
The source is placed at [10m, 30m, 25m] in the 100m*60m*30m searching area.
It can be seen that as the number of UAVs rises, the MSTs will decrease monotonically and the SRs will increase for all four algorithms in Fig. \ref{resultUAV}.
\textit{MUC-OSL} algorithm always outperforms all baselines.
\textit{Adap-PP} has a better performance than \textit{Col-PF} when the number of UAVs is smaller, which indicates that dynamically adjusting flying steps can effectively reduce the localization time.
Then, the MST of \textit{Col-PF} drops faster than \textit{Adap-PP} as the number of UAVs increases, due to the fact that the particle updates of \textit{Col-PF} are more dependent on the cooperation of multiple UAVs.
When the number of UAVs is greater than 5, the MSTs obtained by all algorithms decrease slightly.
As the number of UAVs increases, the SRs of all algorithms increase rapidly and then tend to stay steady when the number of UAVs is greater than 5.

\begin{figure}[ht]
	\begin{center}
		\subfigure[mean search time]{
			\includegraphics[width=4cm]{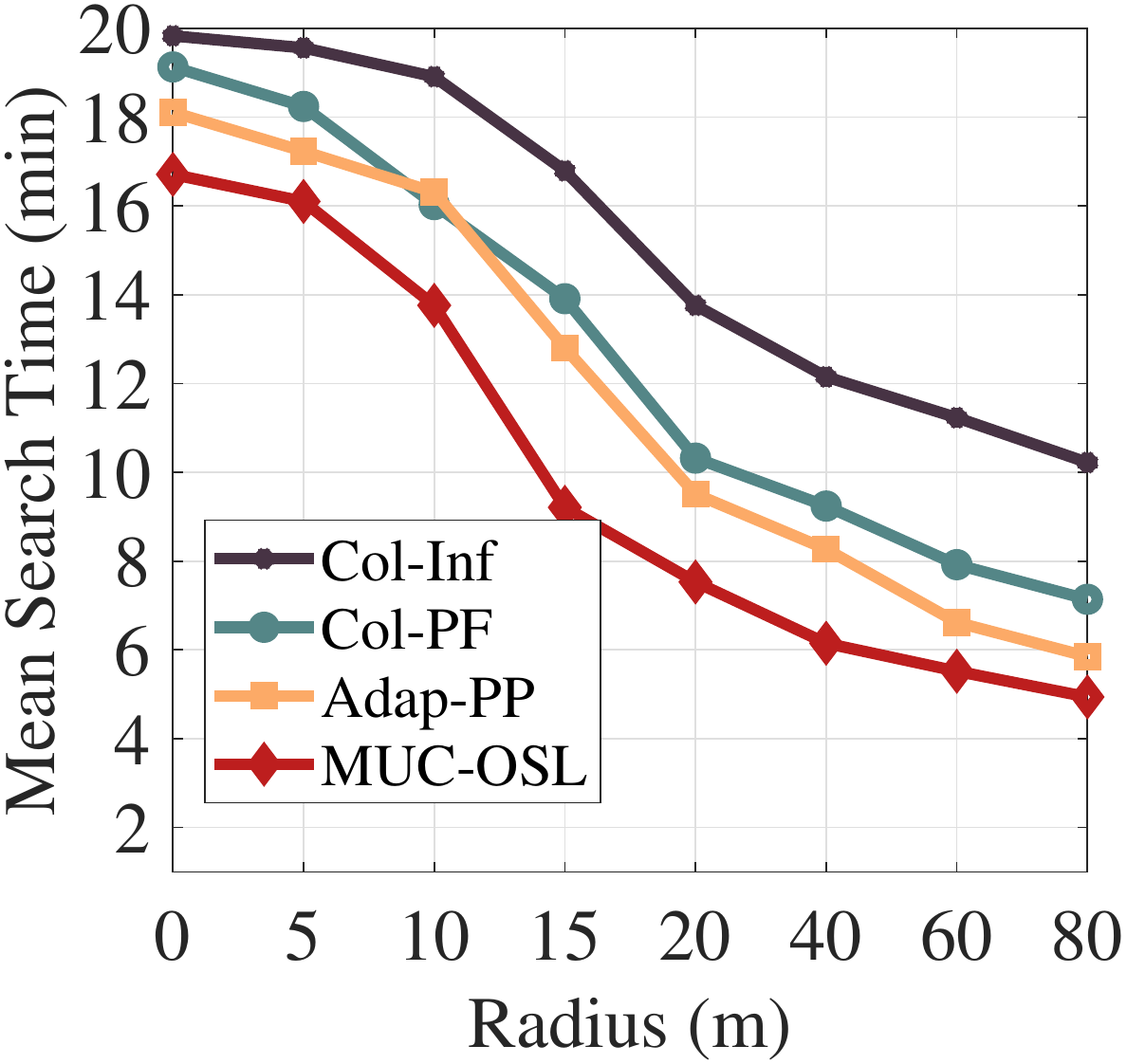}
		}
        \subfigure[success rate]{
			\includegraphics[width=4cm]{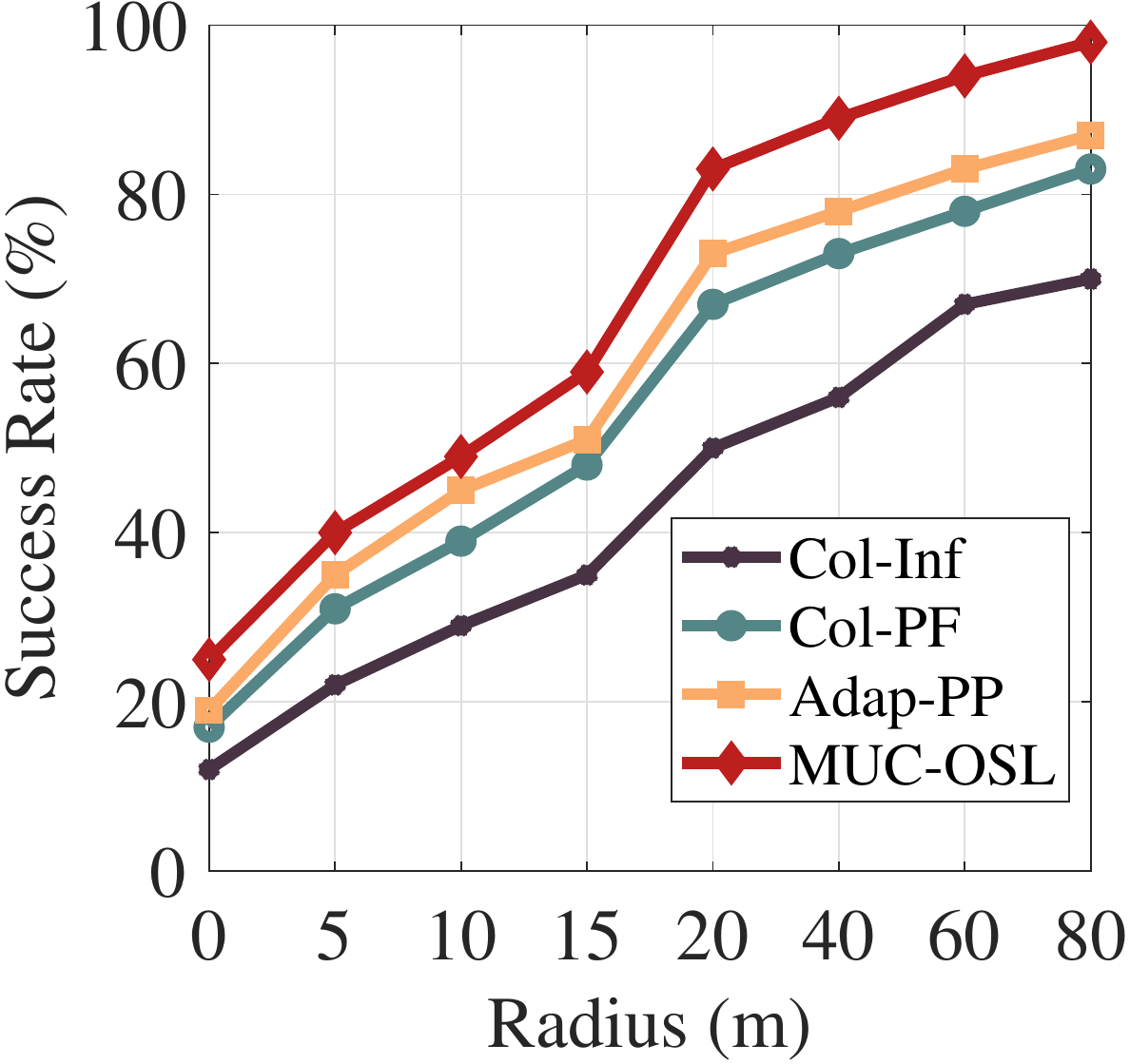}
		}
		\caption{Performance comparison of four algorithms with different communication radii.}
		\label{resultR}
	\end{center}
\end{figure}

\textbf{\textit{3) Different communication radii:}}
As shown in Fig. \ref{resultR}, we observe the performance comparisons with different communication radii, under the assumption that each UAV only shares its local information with neighboring UAVs within the communication radius.
In this simulation scenario, we place 4 UAVs in the 100m*60m*30m area to search odor source.
It corresponds to the non-collaboration case when the communication radius is 0.
The opportunity of multi-UAV collaboration is low when the communication radius is smaller, and thus the performance of \textit{Col-PF} relying on multi-UAV collaboration is poor compared with \textit{Adap-PP}.
As the communication radius increases form 5 to 20, more interactions among neighboring UAVs can be achieved, which can accelerate particle convergence and improve searching accuracy, and thus the performance of \textit{Col-PF} and \textit{MUC-OSL} can be improved significantly.
It can be seen that both MST and SR improve gradually and steady when the communication radius is greater than 20, approaching all-to-all communication in source localization.

\begin{figure}[ht]
	\begin{center}
		\subfigure[mean search time]{
			\includegraphics[width=4cm]{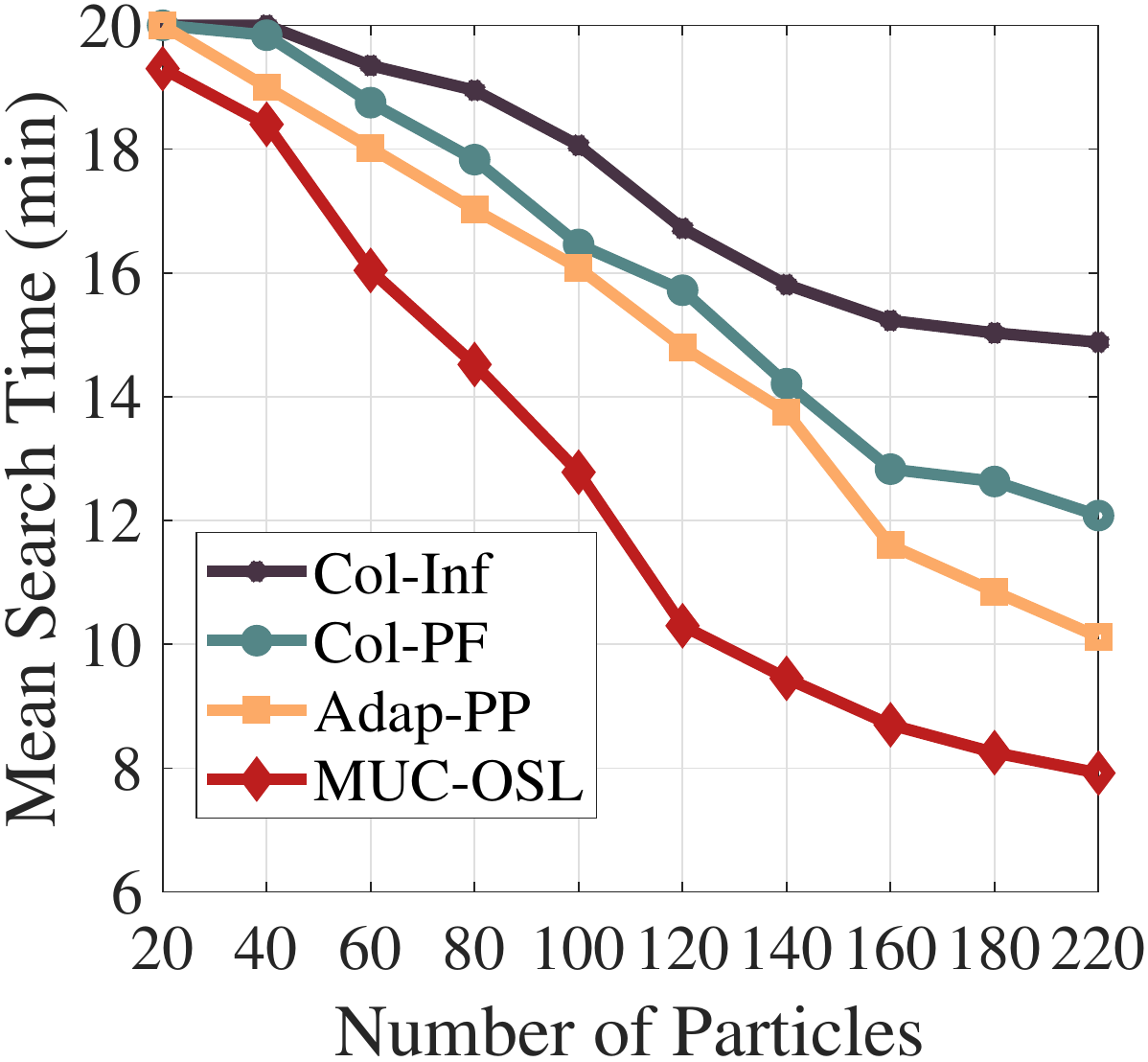}
		}
        \subfigure[success rate]{
			\includegraphics[width=4cm]{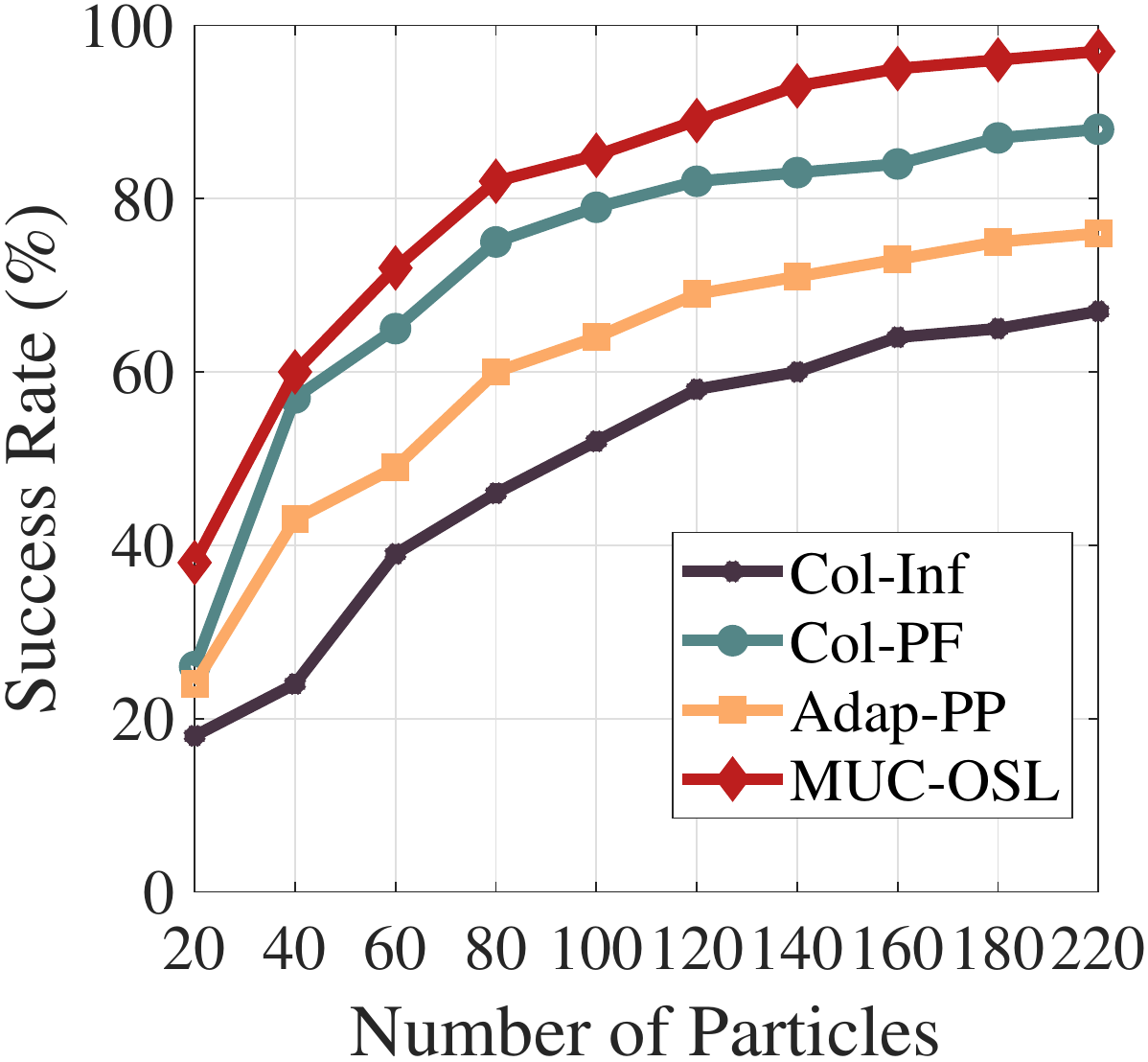}
		}
		\caption{Performance comparison of four algorithms with different numbers of particles.}
		\label{resultP}
	\end{center}
\end{figure}

\textbf{\textit{4) Different particles numbers:}}
The performance comparisons with different numbers of particles are compared as shown in Fig. \ref{resultP}.
We use three UAVs to search in the 100m*60m*30m searching area.
As the number of particles increases from 20 to 220, the MSTs decrease and the SRs increase for all algorithms, indicating that more particles will lead to relatively easier source searching.
However, increasing the number of particles will lead to expensive computation burden.
When utilizing a small number of particles, the MST of \textit{Adap-PP} is high since the estimated source position is inaccurate and then the erroneous path decisions by UAVs lead to path redundancy.
In this condition, \textit{Col-PF} performs better because multi-UAV collaboration enables the position of estimated source to be more accurate.
It can be seen that \textit{MUC-OSL} outperforms \textit{Col-Inf} no matter how many particles there are.
When the particles number is larger than 160, the MST becomes steady and the SR is near 1 for the proposed \textit{MUC-OSL} algorithm.

\subsection{Results for high-fidelity simulator}
\begin{figure*}[ht]
	\begin{center}
		\subfigure[the initial state in GADEN]{
			\includegraphics[width=5cm]{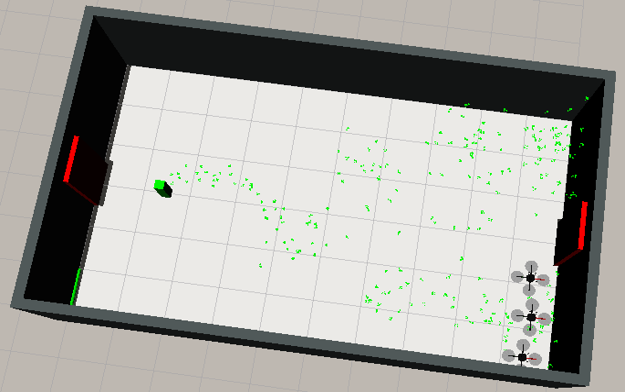}
		}
        \subfigure[a snapshot of the searching process ]{
			\includegraphics[width=5cm]{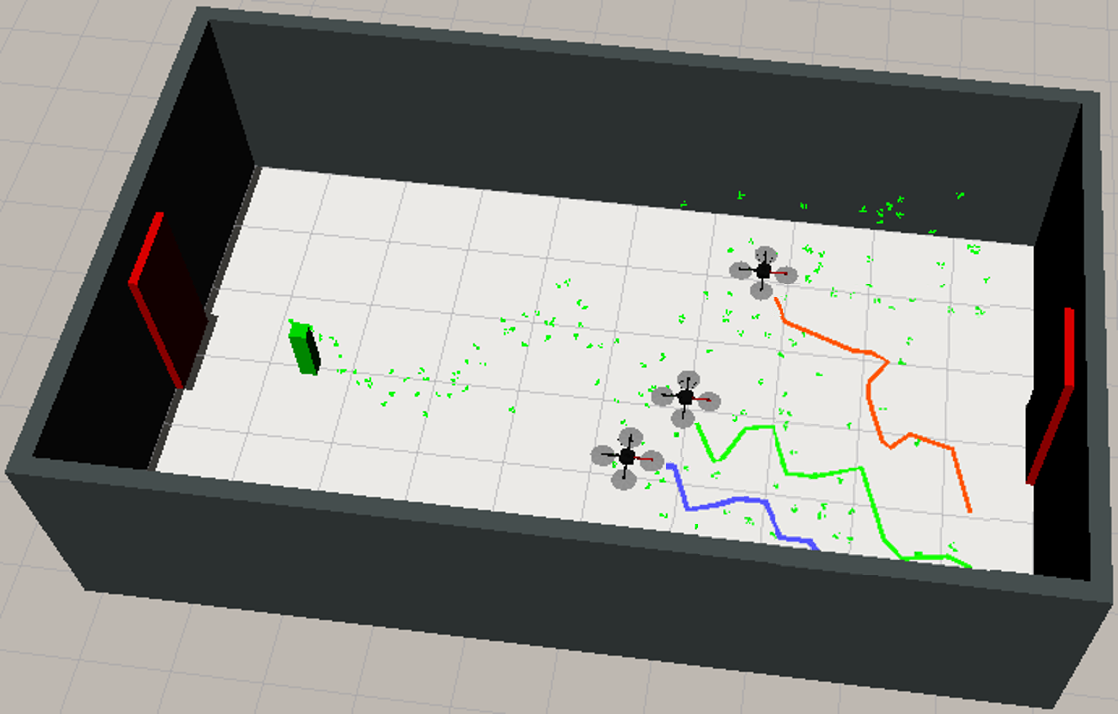}
		}
        \subfigure[the first UAV arrived the source]{
			\includegraphics[width=5cm]{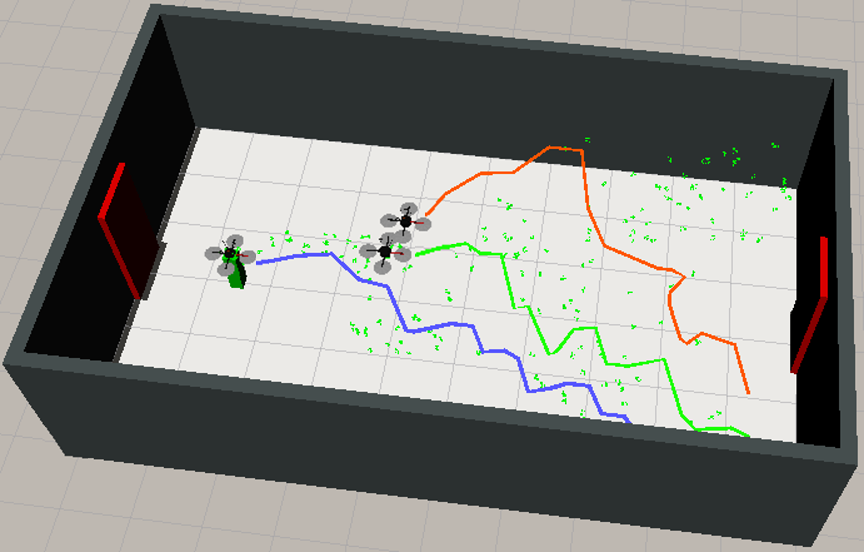}
		}
		\caption{An example of multi-UAV running process in the GADEN simulator.}
		\label{gaden}
	\end{center}
\end{figure*}


\begin{figure*}[htbp]
	\begin{center}
		\subfigure[different searching area sizes]{
			\includegraphics[width=1.6in]{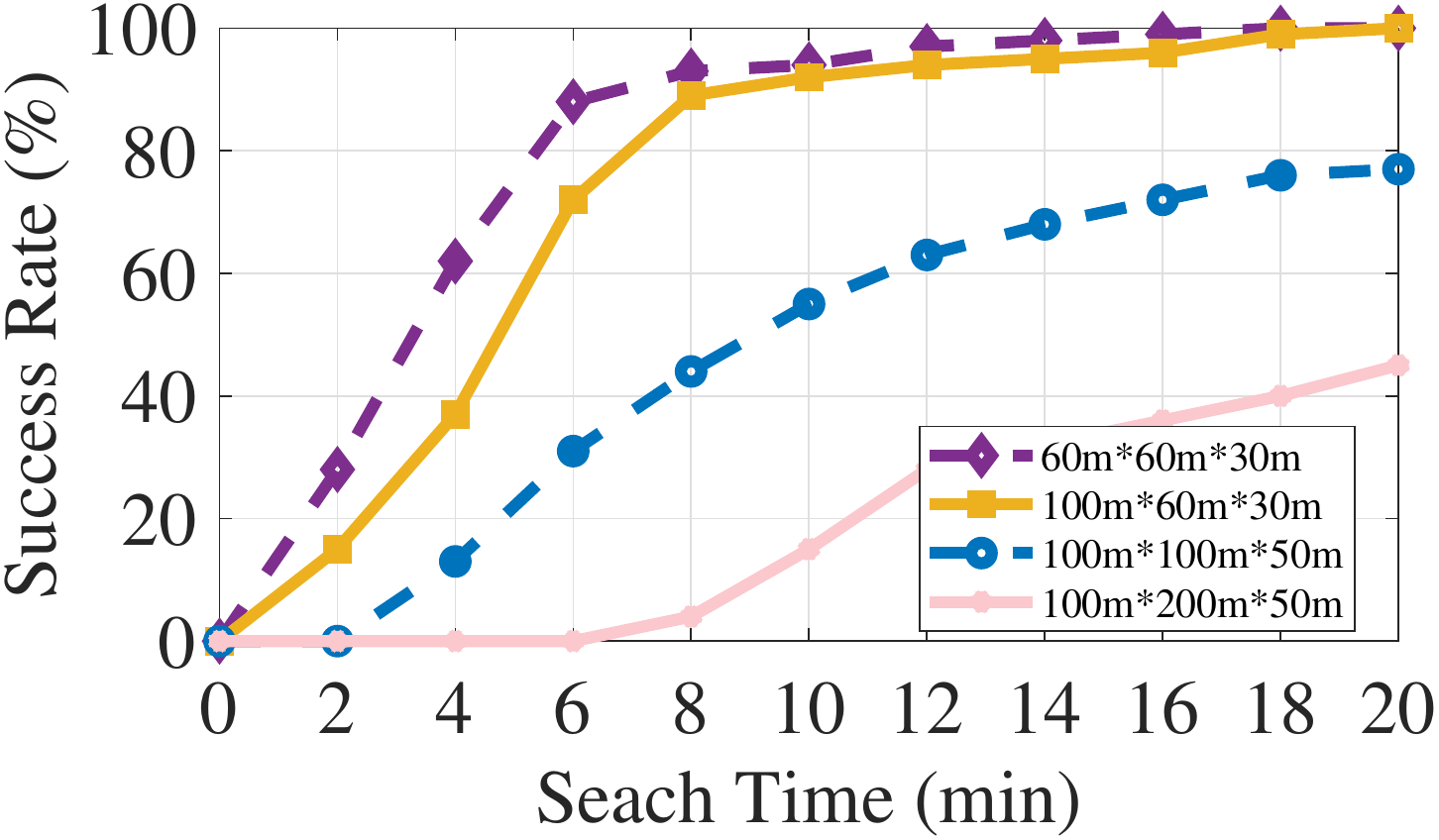}
		}
        \subfigure[different UAVs numbers]{
			\includegraphics[width=1.6in]{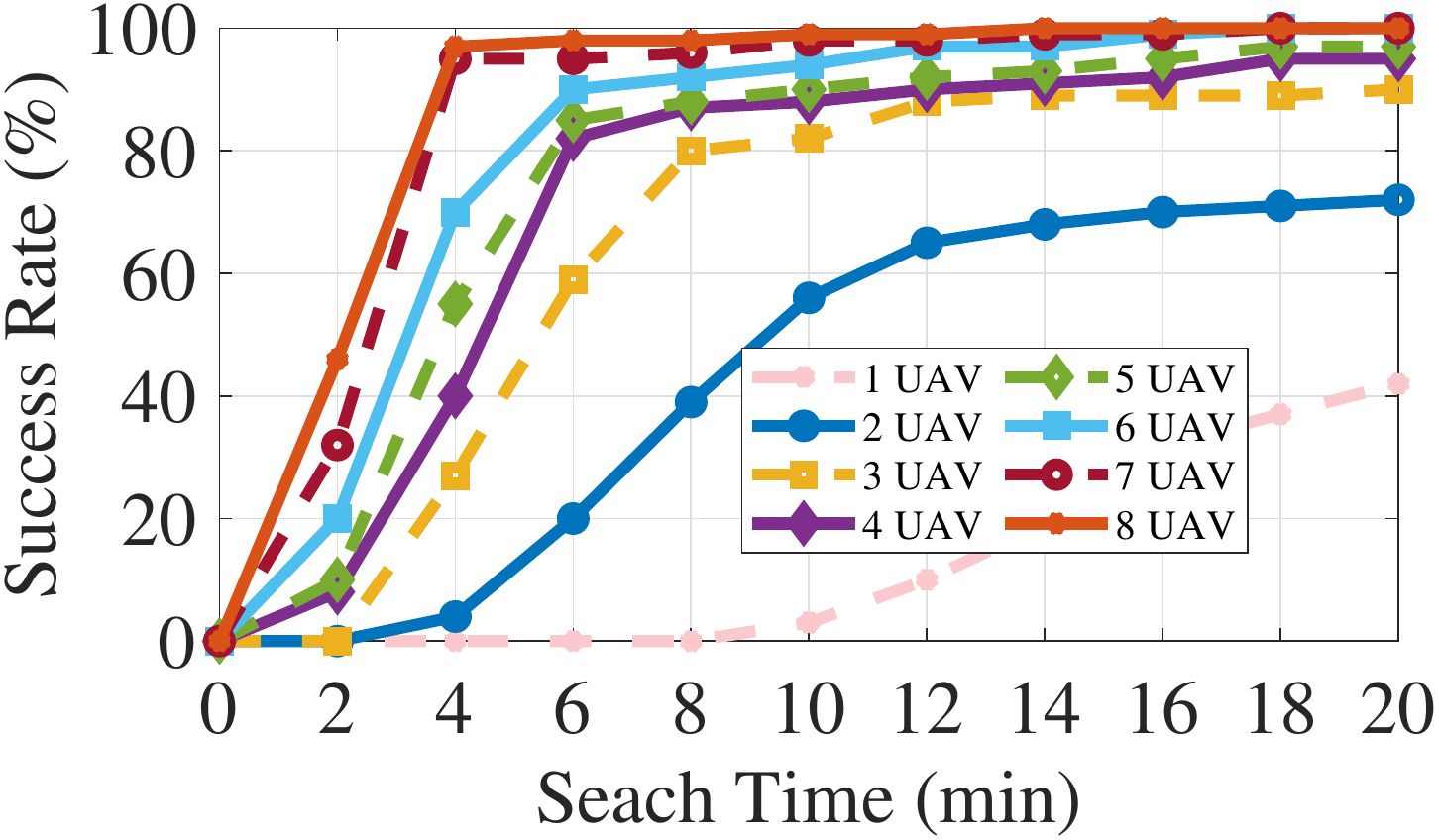}
		}
		\subfigure[different communication radii]{
			\includegraphics[width=1.6in]{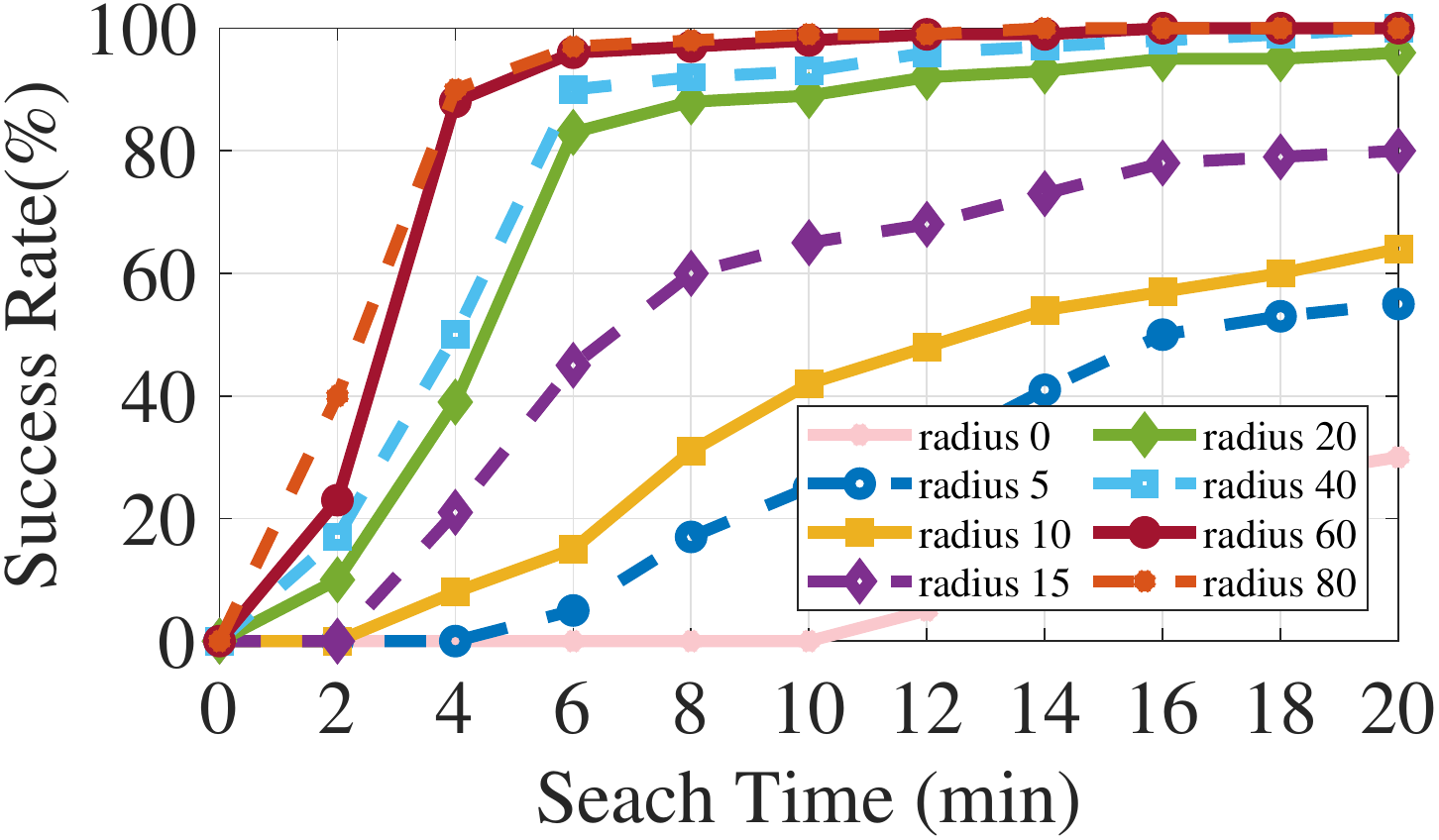}
		}
		\subfigure[different particles numbers]{
			\includegraphics[width=1.6in]{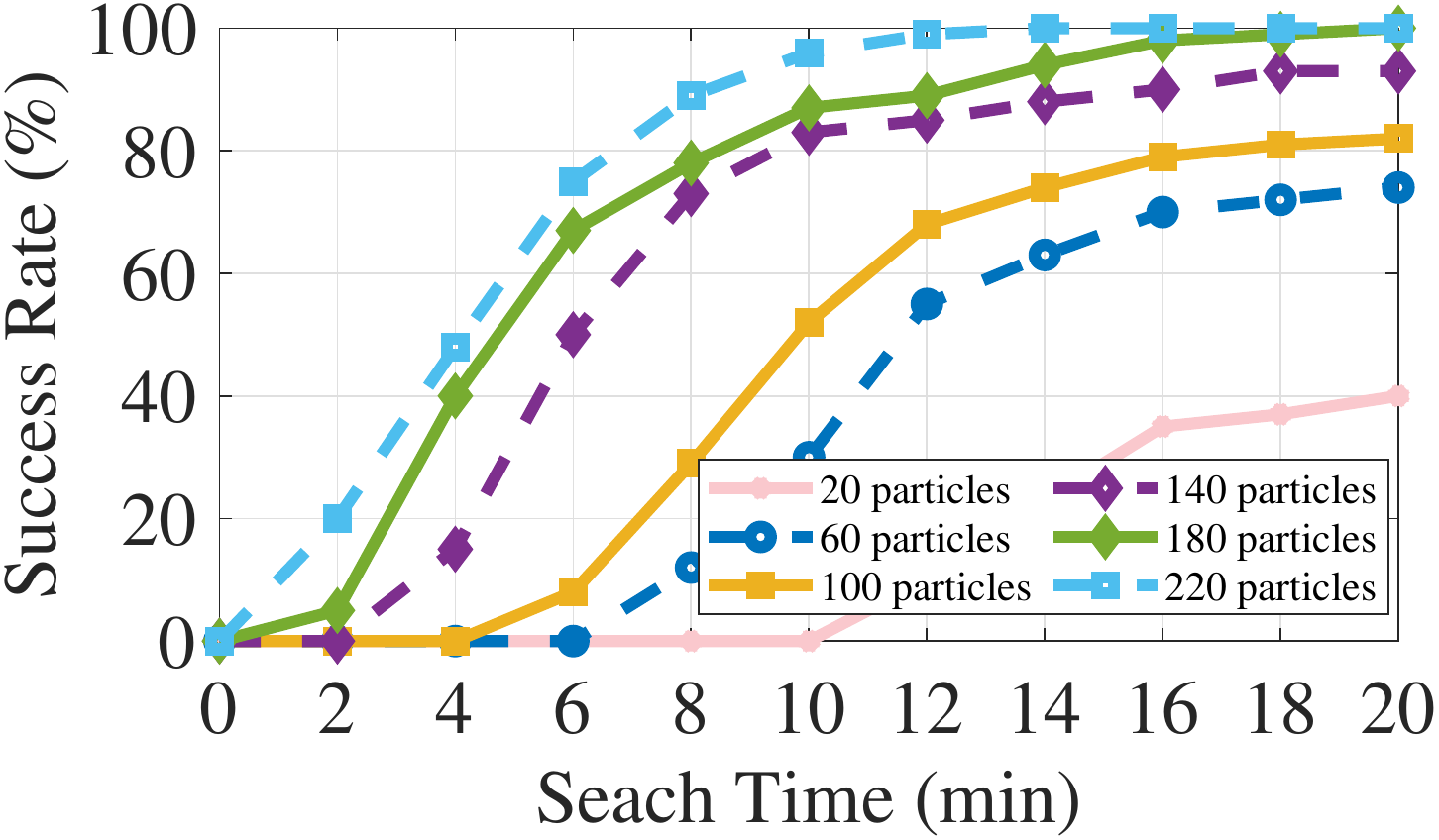}
		}
		\caption{Comparison results in the GADEN simulator.}
		\label{Resultgaden}
	\end{center}
\end{figure*}
The GADEN platform is utilized to verify the algorithm performance in a more realistic plume environment, which models the gas dispersion and wind field in the 3D realistic environment based on computational fluid dynamics and filament dispersion theory \cite{monroy2017gaden}.
We model the UAV using the hector quadrotor package \cite{hectorquadrotor}
 in the GADEN simulator, and each UAV carries anemoscope to measure wind information and photo ionization detector (PID) to capture gas information.
The algorithm parameters settings, including particle parameters, are the same as the previous simulations.
What is more, due to the complexity and unpredictability of the GADEN-generated plume distribution, the Gaussian fitting strategy
is not suitable for the GADEN simulator, which has no influence on the simulation results except for the consumption of communication resources.

Fig. \ref{gaden} shows a running example in the GADEN simulator at different times. 
The gas source,
represented by the green bar, has a certain height and releases gas particulate (small green dots distributed in 3D space) over time.
Three UAVs collaborate to find the gas source, and their travel paths are indicated in blue, green and red, respectively.

The main focus of this study is to investigate the performance of \textit{MUC-OSL} algorithm under a more realistic environment generated by GADEN simulator.
Similar to the previous simulations, the MST and SR of \textit{MUC-OSL} are studied under different parameters.
The comparison results of algorithm performance are represent in Fig. \ref{Resultgaden} under different searching area sizes, varied UAVs numbers, increasing communication radii, and increasing numbers of particles.


Overall, \textit{MUC-OSL} algorithm in GADEN simulator can achieve same or better performance than previous simulations,
which is due to the size of plume diffusion area in the GADEN simulator is larger than the diffusion area size of the rate-based Gaussian plume.
Comparing Fig. \ref{Resultgaden}(a) with Fig. \ref{resultArea},
the similar success rate can be obtained in a shorter time.
For example, in the 100m*60m*30m searching area,
it needs 8 minutes in the GADEN simulator to achieve about 90\% success rate, while it takes more time in the previous simplified simulation.
In Fig. \ref{Resultgaden}(b), it can be seen that more than 80\% success rate can be obtained within 8 minutes when the number of UAVs is greater than 3, which confirms the conclusion of Fig. \ref{resultUAV}.
For all communication radii, Fig. \ref{Resultgaden}(c) achieves a higher success rate over time, compared with Fig. \ref{resultR}. It also reflects that the performance tends to be stable when the communication radius is larger than 20m.
Fig. \ref{Resultgaden}(d) indicates a better performance can be obtained when the number of particles is larger than 140, which is similar to Fig. \ref{resultP}.
Therefore, the results in GADEN simulator also prove the effectiveness of our proposed localization method in a more realistic environment.

\subsection{Resource Consumption Analysis}

\textbf{\textit{1) Movement energy consumption analysis:}}
In the evaluation of UAVs' movement energy consumption in odor source localization, we cite the actually measured energy consumption data in \cite{infocom19csi}, which conducts the real experiments using the DJI Matric 100 to measure three basic components of movement energy model.
The hovering energy consumption is $0.47KJ/s$,
the flying energy consumption is $0.663KJ/s$,
and the energy consumption of turning with different turning angles is $3.415KJ$.
In this paper, we use these data of energy consumption under the same simulation settings.
The whole movement energy consumption of UAVs $E_M$ can be computed using these cited data.

The average energy consumptions of 200 Monte Carlo simulations for the proposed \textit{MUC-OSL} algorithm and \textit{Col-Inf} algorithm are shown in Table \ref{energy2}.
Compared to \textit{Col-Inf} algorithm, \textit{MUC-OSL} algorithm can reduce the movement energy consumption of multiple UAVs up to 8.44\%, which has a great significance especially when searching a wide region.
The reason is that the adaptive path planning strategy in the \textit{MUC-OSL} algorithm can remarkably reduce the number of turning points and hovering points of multiple UAVs with the sacrifice of marginally increasing flying distance, which accounts for the reduction in movement energy consumption.

\begin{table}[ht]
    \caption{Energy consumption comparison.}
    \label{energy2}
\resizebox{\linewidth}{!}
{
\begin{tabular}{ccccc}
\hline
              & hovering time & turning time     & distance & $ E_M$ \\ \hline
\textit{Col-Inf}    & 261s          & 223s         & 893m        &1476.27kJ    \\ \hline
\textit{MUC-OSL}    & 216s          & 190s         & 907m        &1351.71kJ    \\ \hline
\end{tabular}
}
\end{table}

\textbf{\textit{2) Communication and computation analysis:}}
Table \ref{com} shows the comparisons of communication and computation resource consumption between \textit{Col-Inf} and \textit{MUC-OSL}.
Communication energy consumption is quantified by the transmitted data amount, and computation energy consumption is quantified by the number of data bits required to be processed, as described in the energy consumption model in Sect. 3.2. 
Specifically, the communication consumption is represented by $4N$ where $N$ denotes the number of particles in the \textit{Col-Inf} algorithm, since the positions and weights of all particles, i.e., 4 bytes, need to be transmitted among neighboring UAVs.
In our proposed \textit{MUC-OSL} algorithm, the 3D Gaussian fitting method is applied to particles distribution derived by each UAV before exchanging information.
In this way, the transmitted information only includes the feature values of Gaussian distribution, i.e., ${\mu _i^k}$ and $ {\Sigma _i^k}$ for UAV $i$, which greatly reduces the scale of data interaction and saves communication cost.

When the particles number is fixed in the source localization, the computation complexity of particle filter and path planning algorithm in the \textit{Col-Inf} can be denoted as $O(c \cdot N)$ and $O(V_{dir} \cdot (2N +1+ d_{max}))$, and thus the whole computation complexity of \textit{Col-Inf} can be simplified as $O(c \cdot N)$ with a constant $c$.
It can be seen that the number of particles plays a crucial role in computation resource consumption.
In the \textit{MUC-OSL} algorithm, the particles number can be proportionally decreased based on cue-captured frequency and distance gap, and thus the computation complexity can be reduced to $O(c \cdot N')$ where $N'=N/m$ and $m$ is a constant.
Benefitting from the reduction of whole energy consumption, \textit{MUC-OSL} algorithm can be well applied to the source localization of UAVs with limited resources.

\begin{table}[ht]
	\centering
{
\caption{The comparisons of computation and communication cost. }
\label{com}
\small
\begin{tabular}{ccc}
\hline
                  & communication cost     & computation cost \\ \hline
\textit{Col-Inf}         & $4N$                  & $ O(cN)  $     \\ \hline
\textit{MUC-OSL}         &  12                & $O(cN/m)$   \\ \hline
\end{tabular}
}
\end{table}

\section{Conclusion}
In this paper, we have presented a resource-aware source localization algorithm using multiple UAVs in 3D space named \textit{MUC-OSL}, including source estimation based on collaborative particle filter and UAV navigation based on adaptive path planning.
In the estimation phase,
we propose the \textit{Col-PF} algorithm to approximate source probability inference, in order to reduce the communication and computation burden and speed up particle convergence.
In the navigation phase, we design the \textit{Adap-PP} algorithm, where each UAV individually determines its real-time path planning including the next flying direction and the moving steps, aiming to shorten overall search time and improve the efficiency of source localization.
Extensive simulations based on rate-based plume model in a simplified simulator and the realistic plume model generated by GADEN simulator indicate that the proposed \textit{MUC-OSL} algorithm can reduce the mean search time and improve success rate compared to the state-of-the-art.
Moreover, the proposed \textit{MUC-OSL} algorithm can efficiently reduce multi-UAV resource overhead and energy consumption.
In the future, considering the real-world scenarios, it is worth to study the effective source localization in the more complex environment with multiple sources and obstacles.


%
%

%
%

\acknowledgements{This work was supported by the National Natural Science Foundation of China under Grants 62072436, 61732017 and 61872028, and the National Key Research and Development Program of China under Grant 2021YFB2900102.}

\bibliographystyle{cjereport}
\bibliography{mybibfile}

\end{document}